\documentclass[a4paper,twoside,openany,11pt]{memoir} 
\usepackage{minibox}
\usepackage{longtable}
\def\fullheadfoot{0}
\usepackage[english]{babel}
\usepackage{microtype,xspace}
\usepackage[utf8]{inputenc}	

\usepackage{amsfonts,amsmath,amssymb,bm,mathrsfs,mathtools,dsfont} 	
\allowdisplaybreaks 	
\usepackage{xcolor}
\usepackage{hyperref}
\usepackage{slashed}
\usepackage{multicol}

\usepackage{siunitx}
\usepackage{geometry}
\usepackage[sort&compress,numbers,merge]{natbib}
\usepackage{chapterbib}
\addto\captionsenglish{\renewcommand*{\bibname}{References}}
\makeatletter
\renewcommand{\@memb@bchap}{ 
\bibmark \prebibhook
}
\makeatother

\usepackage{graphicx}
\graphicspath{{./Figures/}{./Figures/A_graphs/}{./Figures/Beta_graphs/}{./Figures/T_graphs/}{./Figures/Chp6/}}
\usepackage{caption,subcaption}
\usepackage{multirow}
\renewcommand{\arraystretch}{1.2}
\usepackage{tabularx}
\newcolumntype{Y}{>{\centering\arraybackslash}X}

\usepackage[shortlabels]{enumitem}
\setlist{itemsep=.1em,topsep=.5em}
\SetEnumerateShortLabel{i}{\textit{\roman*}}

\definecolor{red}{rgb}{0.6,.0706,.1373}
\definecolor{blue}{rgb}{0,0.396,0.741}
\colorlet{blueRef}{blue!80!black}
\colorlet{blueLink}{blue!100!black}
\hypersetup{
	colorlinks, 
	bookmarksopen, 
	bookmarksnumbered,
	citecolor=blueLink, 	
	linkcolor=blueLink,	
	urlcolor=blueLink,			
	pdftitle={SuperTracer},    
	pdfsubject={},	
	pdfauthor={},    	
}

\addto\captionsenglish{}

\renewcommand{\contentsname}{Contents}
\renewcommand{\printtoctitle}[1]{}

\settocdepth{subsection}
\newcommand{\toc}{ {
	\hypersetup{linkcolor = black} 
	\vspace*{-.06\textheight}	
	\tableofcontents*
	\thispagestyle{empty} 
} }

\newcommand{\app}[1][Appendices]{
	\renewcommand{\thesubsection}{\Alph{subsection}}
	\numberwithin{equation}{subsection}
	\numberwithin{figure}{subsection}
	\pagestyle{appstyle}
	\sectionlike{#1} 
}

\makeatletter
\newcommand*\ifthispageodd{%
  \checkoddpage
  \ifoddpage
    \expandafter\@firstoftwo
  \else
    \expandafter\@secondoftwo
  \fi
}
\makeatother

\usepackage[noindentafter]{titlesec} 
\counterwithout{section}{chapter} 
\counterwithout{figure}{chapter} 
\counterwithout{table}{chapter} 
\numberwithin{equation}{section} 
\setsecnumdepth{subsubsection}

\DeclareMathAlphabet{\mathsfit}{OT1}{lmss}{m}{sl}
\DeclareMathAlphabet{\mathsfbf}{OT1}{lmss}{bx}{n}
\DeclareMathAlphabet{\mathsfbfit}{OT1}{lmss}{bx}{sl}
 
\DeclareMathVersion{chaptermath}
\SetSymbolFont{operators}{chaptermath}{OT1}{cmbr}{m}{n}
\SetSymbolFont{letters}{chaptermath}{OML}{cmbrm}{b}{it}
\SetSymbolFont{symbols}{chaptermath}{OMS}{cmbrs}{m}{n}
\DeclareMathVersion{subsectionmath}
\SetSymbolFont{operators}{subsectionmath}{OT1}{cmbr}{m}{n}
\SetSymbolFont{letters}{subsectionmath}{OML}{cmbrm}{m}{it}
\SetSymbolFont{symbols}{subsectionmath}{OMS}{cmbrs}{m}{n}

\titleformat{\section}{\centering \Large \bfseries \sffamily \mathversion{chaptermath} \color{blue!90!black} }{\thesection}{15pt}{}{}
\titlespacing{\section}{0pt}{15pt}{5pt}
\titleformat{\subsection}{\large \sffamily \mathversion{subsectionmath} \color{blue!90!black} }{\thesubsection}{10pt}{}{}
\titlespacing{\subsection}{0pt}{10pt}{5pt}
\titleformat{\subsubsection}{\normalsize \sffamily \itshape \mathversion{subsectionmath} \color{blue!80!black} }{\thesubsubsection}{10pt}{}{}
\titlespacing{\subsubsection}{0pt}{10pt}{0pt}

\newcommand{\sectionlike}[1]{\phantomsection \addcontentsline{toc}{section}{#1} \sectionmark{#1}
		\begin{center}
		\needspace{8\baselineskip}
		\Large \bfseries \sffamily \mathversion{chaptermath} \color{blue!90!black} #1  
		\end{center}
	\vspace{-5pt} 
}

\ifnum\fullheadfoot=1
	\makepagestyle{standardstyle}
	\makeoddhead{standardstyle}{\sffamily \mathversion{subsectionmath} \rightmark}{}{\sffamily \bfseries{\thepage}}
	\makeevenhead{standardstyle}{\sffamily \bfseries{\thepage}}{}{\sffamily \mathversion{subsectionmath} \leftmark}
	\makeheadrule{standardstyle}{\textwidth}{\normalrulethickness}
	\makepsmarks{standardstyle}{
		\nouppercaseheads
		\createmark{section}{both}{shownumber}{\ }{\hspace{1.2em}  }
		\createmark{subsection}{right}{shownumber}{}{\hspace{1.2em} }
	}
	
	\copypagestyle{appstyle}{standardstyle}
	\makeevenhead{appstyle}{\sffamily \bfseries{\thepage}}{}{ \sffamily \mathversion{subsectionmath} \leftmark}
	\makepsmarks{appstyle}{
		\nouppercaseheads
		\createmark{section}{both}{nonumber}{\ }{\ \ }
		\createmark{subsection}{right}{shownumber}{}{\ \ }
	}
\else 
	\makepagestyle{standardstyle}
	\makeoddfoot{standardstyle}{}{\sffamily -- {\bfseries\thepage} --}{} 
	\makeevenfoot{standardstyle}{}{\sffamily -- {\bfseries\thepage} --}{}
	\copypagestyle{appstyle}{standardstyle}
\fi

\createplainmark{toc}{both}{\contentsname}
\createplainmark{bib}{both}{\bibname}

\makeatletter
\let\MyIntOrig\int
\def\MyIntSpace{\hspace{-.35em}} 
\def\int{\MyInt}
\def\MyInt{\MyIntOrig\MyIntSkipMaybe}
\def\MyIntSkipMaybe{
	\@ifnextchar_{\MyIntSkipScript}{%
		\@ifnextchar^{\MyIntSkipScript}{%
			\@ifnextchar\limits{\MyIntSkipTok}{%
				\@ifnextchar\nolimits{\MyIntSkipTok}{%
					\MyIntSpace}}}}%
}
\def\MyIntSkipScript#1#2{#1{#2}\MyIntSkipMaybe}
\def\MyIntSkipTok#1{#1\MyIntSkipMaybe}

\newcommand{\pushright}[1]{\ifmeasuring@#1\else\omit\hfill$\displaystyle#1$\fi\ignorespaces}
\makeatother

\usepackage{bbm}
\RequirePackage{fix-cm}
\usepackage{fontawesome}
\usepackage{booktabs}
\usepackage{multirow}
\usepackage{soul}
\usepackage{bbold}
\definecolor{DarkGray}{gray}{0.40}

\newcommand{\tr}{\mathop{\mathrm{tr}} }
\newcommand{\eminus}{\vcenter{\hbox{\scalebox{0.6}[1]{$ - $}}}}	
\newcommand{\ord}[1]{\mathcal{O}\!\left( #1 \right)}

\newcommand{\anticommutator}[2]{\big\lbrace#1, \, #2 \big\rbrace}

\newcommand{\andeq}{\quad \mathrm{and} \quad}

\newcommand{\dd}{\mathop{}\!\mathrm{d}}
\newcommand{\ud}[2]{\phantom{}^{#1}\phantom{}_{#2}}

\makeatletter
\newcommand{\vast}{\bBigg@{3}}
\makeatother

\renewcommand{\L}{\mathcal{L}}

\newcommand{\U}{\mathrm{U}}
\newcommand{\SU}{\mathrm{SU}}

\newcommand{\UV}{\mathrm{UV}}
\newcommand{\EFT}{\mathrm{EFT}}

\newcommand{\go}{$ \gamma_5 $-odd\xspace}

\definecolor{verde}{cmyk}{0.92,0,0.59,0.25}
\newcommand{\pkg}[1]{{\texttt{#1}}\xspace}
\newcommand{\rut}[1]{\texttt{\color{verde} #1}}

\usepackage[utf8]{inputenc}

\usepackage{mmacells}
\mmaSet{
  morefv={gobble=2},
  linklocaluri=mma/symbol/definition:#1,
  morecellgraphics={yoffset=1.9ex}
}

\begin{document}

\thispagestyle{empty}
\renewcommand*{\thefootnote}{\fnsymbol{footnote}}
\begin{center} 
\begin{minipage}{15.5cm}
\vspace{-0.7cm}
\begin{flushright}
{\footnotesize \itshape
MITP-20-076\\
TUM-HEP-1302/20\\[-3pt]
ZU-TH-54/20
}
\end{flushright}

\vspace{-2.1cm}
\includegraphics[width=5cm]{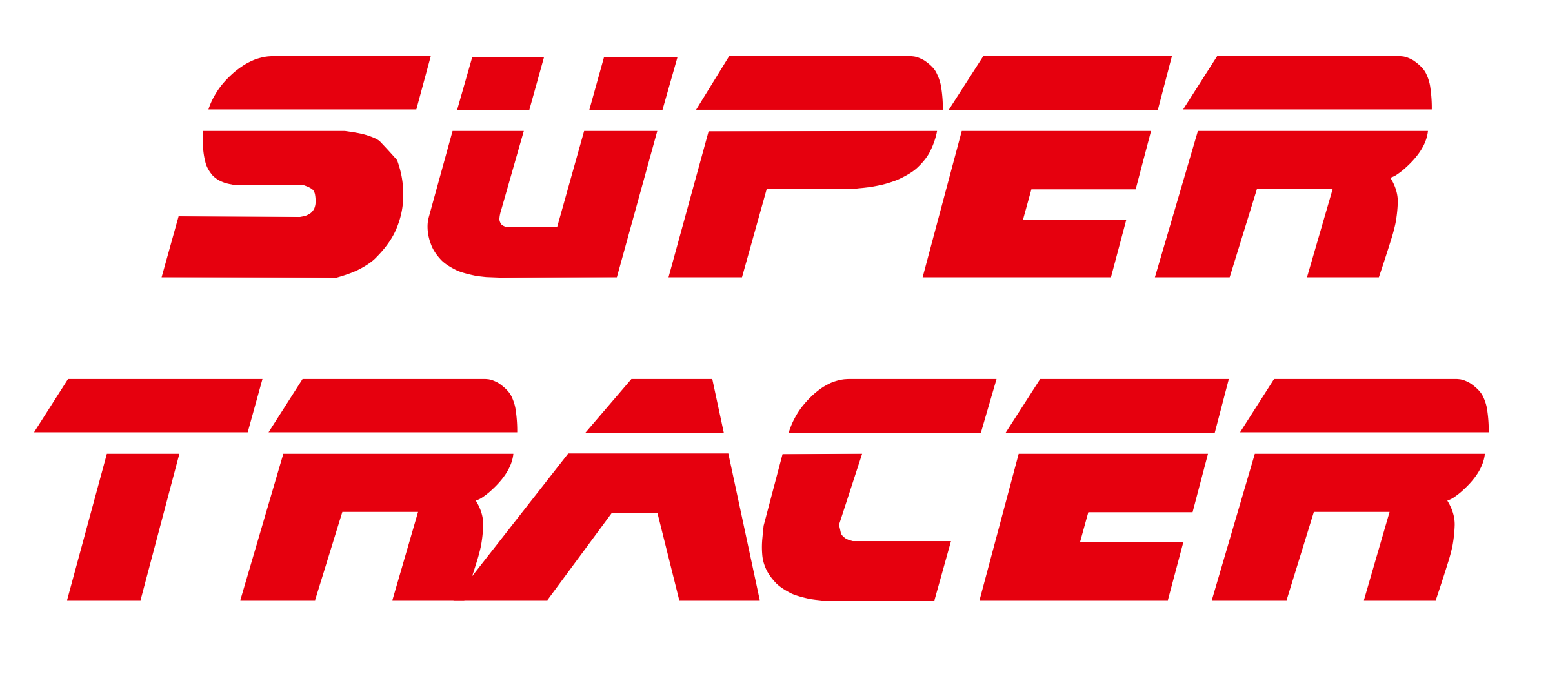}

\vspace{10mm}
\end{minipage}
\end{center}

\begin{center}
    \textcolor{blue!80!black}{\rule{\textwidth}{2pt}}\\
	{\sffamily \bfseries \fontsize{18.}{20}\selectfont \mathversion{chaptermath}
	SuperTracer: A Calculator of Functional Supertraces\\[5pt] for One-Loop EFT Matching}\\[-.5em]
	\textcolor{blue!80!black}{\rule{\textwidth}{2pt}}\\
	\vspace{.05\textheight}
	{\sffamily \mathversion{subsectionmath} \Large Javier Fuentes-Mart\'{\i}n,$^{1}$\footnote{jfuentes@uni-mainz.de} 
	Matthias König,$^{2}$\footnote{matthias.koenig@tum.de}
	Julie Pagès,$^{3}$\footnote{julie.pages@physik.uzh.ch}\\[5pt]
	Anders Eller Thomsen,$^{3,4}$\footnote{thomsen@itp.unibe.ch}
	and Felix Wilsch$^{3}$\footnote{felix.wilsch@physik.uzh.ch}}
	\\[1.25em]
	{ \small \sffamily \mathversion{subsectionmath} 
	$^{1}\,$ PRISMA+ Cluster of Excellence \& Mainz Institute for Theoretical Physics,\\
	Johannes Gutenberg University, D-55099 Mainz, Germany\\[5pt]
	$^{2}\,$ Physik Department T31, Technische Universit\"at M\"unchen,\\
	James-Franck-Str. 1, D-85748 Garching, Germany\\[5pt]
	$^{3}\,$ Physik-Institut, Universit\"at Z\"urich, CH-8057 Z\"urich, Switzerland\\[5pt]
	$^{4}\,$ Albert Einstein Center for Fundamental Physics, Institute for Theoretical Physics,\\ University of Bern, CH-3012 Bern, Switzerland
	}
	\\[.005\textheight]{\itshape \sffamily \today}
	\\[.03\textheight]
\end{center}
\setcounter{footnote}{0}
\renewcommand*{\thefootnote}{\arabic{footnote}}%
\suppressfloats	

\begin{abstract}\vspace{+.01\textheight}
We present \pkg{SuperTracer}, a \pkg{Mathematica} package aimed at facilitating the functional matching procedure for generic UV models. This package automates the most tedious parts of one-loop functional matching computations. Namely, the determination and evaluation of all relevant supertraces, including loop integration and Dirac algebra manipulations. The current version of \pkg{SuperTracer} also contains a limited set of output simplifications. However, a further reduction of the output to a minimal basis using Fierz identities, integration by parts, simplification of Dirac structures, and/or light field redefinitions might still be necessary. The code and example notebooks are publicly available at \href{https://gitlab.com/supertracer/supertracer}{\color{red}\faicon{gitlab}}. 
\end{abstract}

\newpage

\section*{Table of contents}
\toc

\section{Introduction}
\label{sec:intro}

Studying the low-energy phenomenology of a quantum field theory is an important step in most studies in the Standard Model (SM) and beyond. To this end, one constructs the corresponding effective Lagrangian by separating the relevant dynamics from different energy scales and removing the ones lying at high energies. The resulting effective field theory (EFT) can then be used to study the physics at low energies, while keeping large logarithms of the scale hierarchies under control by renormalization-group improvement. A common example of such an approach is when one studies the effects of new-physics (NP) models on flavor observables, where one separates dynamics at and above the weak scale through a series of matching steps from the dynamics at the characteristic scale of the process. In practice, one matches a given UV theory to the Standard Model Effective Field Theory (SMEFT)~\cite{Buchmuller:1985jz,Grzadkowski:2010es} which is then in turn, after renormalization group evolution (RGE) in the SMEFT~\cite{Jenkins:2013zja,Jenkins:2013wua,Alonso:2013hga,Alonso:2014zka}, matched to the Low Energy Effective Theory (LEFT)~\cite{Jenkins:2017jig,Dekens:2019ept,Aebischer:2015fzz} and evolved via the LEFT RG equations~\cite{Jenkins:2017dyc}.

The example of flavor physics also shows the importance of performing the matching steps beyond the leading order, since a great amount of interesting observables (e.g. FCNC processes like rare decays and neutral meson mixing) are generated starting only at one-loop order within the SM. The process of matching NP models to an EFT to study the relevant low-energy phenomenology constitutes a repetitive and time-consuming task, calling for an automated solution. In the recent years, many tools for automated EFT calculations, specially in the context of the SMEFT, have been developed~\cite{Celis:2017hod,Aebischer:2018bkb,Fuentes-Martin:2020zaz,Criado:2017khh,Brivio:2019irc,Gripaios:2018zrz,Criado:2019ugp,Dedes:2019uzs,Hartland:2019bjb,Aebischer:2018iyb,EOS,Straub:2018kue,Brivio:2017btx,Uhlrich:2020ltd,Bakshi:2018ics}. In particular, tools for RGE in the SMEFT and LEFT and one-loop matching of the SMEFT to the LEFT~\cite{Celis:2017hod,Aebischer:2018bkb,Fuentes-Martin:2020zaz}, tree-level EFT matching of generic UV models~\cite{Criado:2017khh} (see also~\cite{deBlas:2017xtg}), as well as partial one-loop EFT matching results~\cite{Bakshi:2018ics,Kramer:2019fwz,Angelescu:2020yzf,Ellis:2020ivx} are available. Moreover, the \pkg{Matchmaker} package (not yet released) will automate the diagrammatic EFT matching of generic UV models~\cite{Matchmaker,Brivio:2019irc}.
However, to our knowledge, no tool for complete one-loop EFT matching is currently publicly available. We provide here a first building block in this direction by introducing  \pkg{SuperTracer}, a \pkg{Mathematica} package aimed at facilitating the one-loop EFT matching of generic UV models using path integral methods.

The path integral formulation of one-loop EFT matching~\cite{Gaillard:1985uh,Chan:1986jq,Cheyette:1987qz,Chan:1985ny,Fraser:1984zb,Aitchison:1984ys,Aitchison:1985pp,Aitchison:1985hu,Cheyette:1985ue,Chan:1986jq,Dittmaier:1995cr,Dittmaier:1995ee,Henning:2014wua,Drozd:2015rsp,delAguila:2016zcb,Boggia:2016asg,Henning:2016lyp,Ellis:2016enq,Fuentes-Martin:2016uol,Zhang:2016pja,Ellis:2017jns,Summ:2018oko,Kramer:2019fwz,Cohen:2019btp,Angelescu:2020yzf,Ellis:2020ivx,Cohen:2020fcu} has clear advantages over the diagrammatic procedure. For example, one does not need to handle Feynman diagrams nor symmetry factors, and one obtains directly the complete set of EFT operators together with their matching coefficients, without requiring any prior knowledge of the EFT operator structure, symmetries, etc. All of these points and the systematic nature of the procedure render the functional approach exceptionally suited to be implemented in a computer program. As we describe in detail in Section~\ref{sec:method}, at the heart of the functional one-loop matching procedure is the evaluation of functional supertraces. \pkg{SuperTracer} provides the full list of relevant supertraces for a given set of interactions and automates their calculation to an arbitrary order in the heavy mass expansion, limited only by the rapidly increasing complexity of the calculation at higher orders.\footnote{During the completion of this project, we became aware of \pkg{STrEAM}~\cite{Cohen:2020qvb}, a package with a similar scope that is released at the same time.} Furthermore, it provides the option of inputting substitutions for the interactions in terms of fields, and applies some output simplifications. These arguably constitute the most tedious parts of one-loop functional matching computations. An important difference with other functional approaches is that the evaluation of the supertraces is performed in a manifestly covariant way by the application of the so-called Covariant Derivative Expansion (CDE)~\cite{Gaillard:1985uh,Chan:1986jq,Cheyette:1987qz}. 

The structure of this paper is as follows: In Section~\ref{sec:method}, we outline the functional procedure used in \pkg{SuperTracer}. Readers unconcerned about the theoretical details can skip to Section~\ref{subsec:summaryPI}, where we list all the steps that are needed to perform the one-loop EFT matching. Section~\ref{sec:SuperTracer} then gives a short manual of the package and its functions. In Section~\ref{sec:examples} we demonstrate the usage of the package with two examples: a toy model with a heavy vector-like fermion and an $S_1$ scalar leptoquark extension of the SM. We conclude in Section~\ref{sec:conclusions}. Further details about \pkg{SuperTracer} special variables and crosschecks are given in two appendices.

\section{The functional matching procedure}
\label{sec:method}

Consider a general theory $\L_\UV[\eta_H,\eta_L]$, whose field content can be split into heavy $\eta_H$ and light $\eta_L$ degrees of freedom, satisfying $m_H\gg m_L$. Our aim is to compute the EFT resulting from integrating out the heavy degrees of freedom $\eta_H$ at the one-loop order. This can be done following a path integral approach for the effective action of the theory. Collecting heavy and light fields into a field multiplet of the form $\eta = (\eta_H\;\eta_L)^\intercal$,\footnote{For charged degrees of freedom, the field and its complex conjugate enter as separate components in $\eta$, as we describe in more detail in~\ref{subsec:summaryPI}.} the fields are split into background-field configurations $ \hat{\eta} $ satisfying the EOMs and quantum fluctuations $ \eta $, i.e. we let $\eta\to \hat\eta+\eta$. The effective action of the theory is then given by the path integral
\begin{align}
e^{i\,\Gamma_\UV[\hat{\eta}]}= \int\,\mathcal{D}\eta\;\exp\! \left(i\! \int \dd^dx\;\L_\UV[\eta+ \hat{\eta}]\right)\,.
\end{align}
Diagrammatically, the background part corresponds to tree-level lines in Feynman graphs, while lines inside loops arise from the quantum fields. Therefore, at the one-loop level, one only needs to consider terms with up to two quantum fields, since terms with more only produce contributions at higher loop orders. 
The Lagrangian expansion up to terms quadratic in $\eta$ reads
\begin{align}
\L_\UV[\hat\eta+\eta]=\L_\UV[\hat\eta]+\frac{1}{2}\,\bar\eta_i\left.\frac{\delta^2\L_\UV}{\delta\eta_j\delta\bar\eta_i}\right|_{\eta=\hat\eta}\,\eta_j+\mathcal{O}(\eta^3)\,,
\end{align}
where the bar denotes the conjugate and $\delta/\delta\eta_i$ is the functional derivative with respect to $\eta_i$. The first term $\L_\UV[\hat\eta]$ depends only on the classical field configurations and yields the tree-level effective action. At energies much lower than the mass of the heavy fields, one can perform a local expansion in inverse powers of $m_H$ of the heavy field EOMs to eliminate $\hat{\eta}_H$ in favor of the light fields. As a result, we obtain the tree-level EFT, namely we have $\L_\UV[\hat\eta_L,\hat \eta_H(\hat \eta_L)]=\L_\EFT^{\scriptscriptstyle (0)}$, with $\L_\EFT^{\scriptscriptstyle (0)}$ being the tree-level EFT Lagrangian. The linear term in the expansion of $\L_\UV$ around the background fields is, up to a total derivative, proportional to the EOMs evaluated at $\eta=\hat\eta$ and thus vanishes. From the quadratic piece, we identify the fluctuation operator, with the generic form
\begin{align}\label{eq:flucOp}
\mathcal{O}_{ij}\equiv\left.\frac{\delta^2\L_\UV}{\delta\eta_j\delta\bar\eta_i}\right|_{\eta=\hat\eta} &= \delta_{ij}\,  \Delta_i^{\eminus1} - X_{ij}\,,
\end{align}
with $\Delta_i^{\eminus1}$ being the inverse propagator of $\eta_i$ given as\footnote{The vector propagator is gauge dependent, but we work exclusively in the Feynman gauge for the quantum fluctuations as a matter of practicality. This does not imply any particular choice for the classical gauge fields, which remain in the general $R_\xi$ gauge. See~\cite{Henning:2014wua} for elaboration on the subject of heavy vectors in the functional method.}
\begin{align}\label{eq:Prop}
\Delta_i^{\eminus1}&=\left\{
\begin{array}{ccc}
P^2-M_i^2 & \qquad\quad & \mathrm{(scalar)}\\
\slashed{P}-M_i && \mathrm{(fermion)}\\
-g^{\mu\nu}(P^2-M_i^2) && \mathrm{(vector)}\\
\end{array}
\right.\,,
\end{align}
where $P_\mu$ is the Hermitian covariant derivative operator $P_\mu(\hat x,\hat q)=\hat q_\mu+g_G \,G^a_\mu(\hat x)\, T^a$, with $\hat q_\mu=i\partial_\mu$ and $P_\mu=i D_\mu$ in position space, while the $X$ terms encode the particle interactions. For practical purposes, we consider the mass operators of the light fields as perturbative interaction terms in $X$ rather than part of the free Lagrangian, so their Feynman propagators appear as the ones of fully massless particles. Namely, we take $M_i=0$ in~\eqref{eq:Prop} for the light fields.

The one-loop effective action, thus, reads
\begin{align}
e^{i\Gamma_\UV^{\scriptscriptstyle (1)}} = \int\,\mathcal{D}\eta\;\exp\! \left(i\! \int \dd^dx \,\frac{1}{2}\;\bar\eta\,\mathcal{O}\,\eta\right)\,.
\end{align}
This is a Gaussian path integral whose functional integration yields
\begin{align}
e^{i\,\Gamma_\UV^{\scriptscriptstyle (1)}}=(\mathrm{SDet}\,\mathcal{O})^{\eminus \frac{1}{2}}\Longrightarrow \Gamma_\UV^{\scriptscriptstyle (1)}=\frac{i}{2}\;\mathrm{STr}\,\ln\,\mathcal{O}\,,
\end{align}
where the superdeterminant $\mathrm{SDet}$ is a generalization of the regular determinant to the case of supermatrices, i.e. matrices with Grassmann (fermionic) and ordinary (bosonic) entries. Similarly, the supertrace $\mathrm{STr}$ is a generalization of the trace to the case of supermatrices, carrying opposite signs for fermionic and bosonic degrees of freedom. Using the property $\mathrm{STr} \ln (A \,B) = \mathrm{STr}\,\ln A + \mathrm{STr}\,\ln B$, valid even for non-commuting operators, and the form of $\mathcal{O}$ in~\eqref{eq:flucOp}, we get
    \begin{align}\label{eq:protoMaster}
    \Gamma_\UV^{\scriptscriptstyle (1)}=\frac{i}{2}\,\mathrm{STr}\,\ln\,\Delta^{\eminus1}+\frac{i}{2}\,\mathrm{STr}\,\ln\,(1-\Delta X)\,.
    \end{align}
This equation provides the essential building blocks for determining the one-loop EFT. However, $\Gamma_\UV$ contains all possible loop contributions, including those that would correspond to one-loop matrix elements with the tree-level EFT Lagrangian. A crucial simplification takes place by splitting $\Gamma_\UV^{\scriptscriptstyle (1)}$ into \textit{hard}- and \textit{soft}-momentum regions using the so-called method of ``expansion by regions"~\cite{Beneke:1997zp,Jantzen:2011nz},
\begin{align}
\Gamma_\UV^{\scriptscriptstyle (1)}=\left.\Gamma_\UV^{\scriptscriptstyle (1)}\right|_\mathrm{hard}+\left.\Gamma_\UV^{\scriptscriptstyle (1)}\right|_\mathrm{soft}\,,
\end{align}
and identifying the one-loop EFT Lagrangian with the hard part of the effective action of the UV theory~\cite{Fuentes-Martin:2016uol,Zhang:2016pja}:
\begin{align}
\left.\Gamma_\UV^{\scriptscriptstyle (1)}\right|_\mathrm{hard}= \! \int\,\dd^dx\;\L_\EFT^{\scriptscriptstyle (1)}\,.
\end{align}
More precisely, contributions from the hard region directly correspond to those encoded in the short-distance EFT Wilson coefficients (WCs) in $\L_\EFT^{\scriptscriptstyle (1)}$, while contributions from the soft region are the same as those from the long-distance EFT matrix elements with $\L_\EFT^{\scriptscriptstyle (0)}$. The loops containing heavy particles yield contributions from the region of hard loop momenta $p\sim m_H$, and from the soft momentum region, $p\sim q_i, m_L$ with $q_i$ being any light-particle external momenta satisfying $q_i \ll m_H$. On the other hand, loops of light particles receive contributions only from the soft momentum region. The method of expansion by regions states that the contribution of each region is obtained in dimensional regularization by expanding the loop integrand into a Taylor series in the parameters that are small there and then integrating every region over the full $d$-dimensional space of loop momenta. This statement holds up to a mismatch of divergences. Identifying the hard region with the WCs would render them infrared divergent. The mismatch is resolved once one also includes the hard region of the EFT amplitudes, which are all proportional to the scaleless integral
\begin{align}
    \int \frac{\dd^dp}{(2\pi)^d}\,\frac{1}{p^4} \propto \frac{i}{16\pi^2}\left( \frac{1}{\epsilon_{\scriptscriptstyle\UV}}-\frac{1}{\epsilon_{\scriptscriptstyle\mathrm{IR}}}\right)\,,
\end{align}
and have to be subtracted from the hard part, exchanging all IR divergences with UV ones. In practice, one simply does not differentiate between $\epsilon_{\scriptscriptstyle\UV}$ and $\epsilon_{\scriptscriptstyle\mathrm{IR}}$ and skips this last step. The trade-off is that it becomes less transparent whether the scale dependences in the matching coefficients are related to the renormalization of the UV theory or the EFT, unless one explicitly computes the counterterms of the UV theory.

Since $\Delta X\sim m_H^{\eminus 1}$ in the hard region,\footnote{In the fermionic case, we have $\Delta\sim p^{\eminus1}\sim m_H^{\eminus1}$, while $X$ can be at most of $\mathcal{O}(1)$. On the other hand, in the bosonic case, $\Delta\sim p^{\eminus2}\sim m_H^{\eminus2}$ while the interactions can be at most of $\mathcal{O}(m_H)$. Note that this counting holds even if $\L_\UV$ is itself an EFT, since the EFT validity requires $p,m_H\ll\Lambda$ for $\Lambda$ being the EFT cut off.} we can Taylor expand the second logarithm in~\eqref{eq:protoMaster} yielding the master formula for one-loop EFT matching~\cite{Cohen:2020fcu}:
    \begin{align}\label{eq:Master}
    \int\,\dd^dx\;\L_\EFT^{\scriptscriptstyle (1)}=\frac{i}{2}\,\mathrm{STr}\,\ln\,\Delta^{\eminus1}\Big|_\mathrm{hard}-\frac{i}{2}\sum_{n=1}^\infty\frac{1}{n}\,\mathrm{STr}\big[(\Delta X)^n\big]\Big|_\mathrm{hard}\,.
    \end{align}
This formula provides the EFT Lagrangian in terms of two types of terms: \textit{log-type} and \textit{power-type} supertraces. As can be seen, the log-type supertrace only depends on the heavy particle propagators,\footnote{Note that log-type traces with light-field propagators do not contain any heavy scales and, hence, only produce soft contributions.} and is therefore universal. Namely, it only depends on the heavy particles present in the theory, but not on their interactions. On the other hand, the power-type terms depend on the particle interactions, both heavy and light, encoded in $X$. Since, as we mentioned before, $\Delta X$ is at most of $\mathcal{O}(m_H^{\eminus 1})$ in the hard momentum expansion, this provides a natural truncation of the series in terms of the EFT expansion in inverse powers of $m_H$.

\subsection{Covariant evaluation of supertraces} \label{sec:CDE}

The operators appearing in the functional supertraces needed for one-loop matching are of the form $Q(P_\mu,U_k(\hat x))$, having a well-defined rational expansion in its arguments, where $P_\mu$ is the covariant derivative operator defined in the previous section and $U_k$ are a set of momentum-independent functions. The supertrace acting on $Q$, which includes also the trace in momentum space, is given by
\begin{align}
\mathrm{STr}\, Q(P_\mu,U_k)&= \pm\! \int \frac{\dd^dp}{(2\pi)^d}\,\langle p|\,\mathrm{tr}\,Q(P_\mu,U_k)\,|p\rangle\,,
\end{align}
where $+$ ($-$) is for bosonic (fermionic) degrees of freedom, and $\mathrm{tr}$ denotes the trace only over internal degrees of freedom, e.g. gauge, spin, flavor, etc. It is convenient to use the completeness relation of position states, $\int\, \dd^dx\, | x\rangle\langle x|=\mathbb{1}$, to express $Q$ in position space:
\begin{align}\label{eq:preCDE}
\mathrm{STr}\, Q(P_\mu,U_k)&= \pm\! \int \dd^dx\int \frac{\dd^dp}{(2\pi)^d}\,e^{ipx}\,\mathrm{tr}\, Q(P_\mu,U_k(x))\,e^{\eminus ipx}\nonumber \\
&= \pm\! \int \dd^dx\int \frac{\dd^dp}{(2\pi)^d}\,\mathrm{tr}\, Q(P_\mu + p_\mu,U_k(x))\,.
\end{align}
In its current form, this expression is not manifestly covariant. At this point, it is useful to apply a path integral transformation, the so-called CDE expansion~\cite{Gaillard:1985uh,Chan:1986jq,Cheyette:1987qz}, that makes this expression manifestly covariant by putting all instances of $P_\mu$ into commutators of the form $[P_\mu,P_\nu]$, $[P_\mu,[P_\nu,P_\rho]]$, $[P_\mu,U_k]$, etc. The CDE expansion consists in sandwiching the expression in~\eqref{eq:preCDE} between the operators $e^{\eminus P\cdot\partial_p}$ and $e^{P\cdot\partial_p}$:
\begin{align}
\mathrm{STr}\, Q(P_\mu,U_k)&= \pm\! \int \dd^dx\int \frac{\dd^dp}{(2\pi)^d}\,e^{ \eminus P\cdot\partial_p}\,\mathrm{tr}\, Q(P_\mu + p_\mu,U_k(x))\,e^{P\cdot\partial_p}\,,
\end{align}
where $\partial_p$ denotes the partial derivative with respect to the loop momentum $p$. The operator $e^{P\cdot\partial_p}$ is trivially unity when acting to the right, while the operator $e^{\eminus P\cdot\partial_p}$ also becomes unity when it is made to act from the left due to the vanishing of total derivatives under integration, so the supertrace remains invariant under this operation.\footnote{This invariance does not rely on the cyclic property of the trace, which has already been evaluated for momentum coordinates.} However, when passing $e^{\eminus P\cdot\partial_p}$ through $Q$ to cancel against $e^{P\cdot\partial_p}$, it has the desired effect of putting all $P$'s into commutators. More precisely, this transformation yields 
\begin{align}
e^{\eminus P\cdot\partial_p}\,(p_\mu + P_\mu)\, e^{P\cdot\partial_p} &=p_\mu + i\,\tilde{G}_{\mu\nu} \,\partial_p^\nu\,,\nonumber\\
\tilde{U}_k\equiv e^{\eminus P\cdot \partial_p} \,U_k\,e^{P\cdot\partial_p}&= \sum_{n=0}^{\infty} \dfrac{(-i)^n }{n!}\, (D_{\{\alpha_1,\ldots \alpha_n\}}  U_k) \,\partial^{\alpha_1}_p \cdots \partial_p^{\alpha_n},
\end{align}
where the parenthesis denotes that the derivatives act in commutators as per usual, e.g. $(D_\mu A)\equiv[D_\mu,A]$, $(D_\mu D_\nu\, A)\equiv[D_\mu,[D_\nu,A]]$, etc., and 
    \begin{align} \label{eq:G_tilde}
    \tilde{G}_{\mu \nu} \equiv \sum_{n=0}^{\infty} \dfrac{(-i)^n }{(n+2) n!}\, (D_{ \{\alpha_1,\ldots \alpha_n\} }  G_{\mu \nu})\, \partial_p^{\alpha_1} \cdots \partial_p^{\alpha_n}\,,
    \qquad D_{\{\mu_1,\cdots\mu_n\}}\equiv \frac{1}{n!} \sum_{\sigma\in \mathcal{S}_n}D_{\mu_{\sigma(1)}}\cdots D_{\mu_{\sigma(n)}}\,.
    \end{align}
Since $Q(P_\mu,U_k(\hat x))$ has a well-defined rational expansion in its arguments this implies 
\begin{align}\label{eq:covariantSTr}
\mathrm{STr}\, Q(P_\mu,U_k)&= \pm \int \dd^dx\int \frac{\dd^dp}{(2\pi)^d}\,\mathrm{tr}\, Q\big(p_\mu + i\,\tilde{G}_{\mu\nu} \,\partial_p^\nu,\tilde U_k(x)\big)\,,
\end{align}
yielding the desired manifestly covariant expression for the supertrace of $Q$.

\subsection{Explicit evaluation of the relevant supertraces} 

In this section, we outline how to apply the covariant method to the log- and power-type supertraces.

\subsubsection{Log-type supertraces}

For the log-type supertraces, we have to evaluate $\mathrm{STr}\ln\Delta_{\eta_H}^{\eminus1}\big|_\mathrm{hard}$ for all possible $\eta_H$ propagators defined in~\eqref{eq:Prop}. To apply the covariant supertrace evaluation in~\eqref{eq:covariantSTr}, we need to show first that $Q(P_\mu)=\ln\Delta_{\eta_H}^{\eminus1}(P_\mu)$ satisfies the requirement of having a well-defined expansion in $P_\mu$. This can be shown by writing an integral representation of the logarithm: 
\begin{align}
\ln\Delta_{\eta_H}^{\eminus1}=\int_z^1 \dd\xi\,\frac{\dd(\Delta_{\eta_H}^\xi)^{\eminus1}}{\dd\xi}\,\Delta_{\eta_H}^\xi- \ln \Delta_{\eta_H}^z \,,
\end{align}
with $\Delta_i^\xi$ defined as $\Delta_i$ in~\eqref{eq:Prop} but replacing $M_i$ by $\xi M_i$, such that $\dd\big(\Delta_{\eta_H}^\xi\big)^{\eminus1}\!/\!\dd\xi$ does not depend on $p$. By taking the $z\to\infty$ limit, it is clear that $Q=\ln\Delta_{\eta_H}^{\eminus1}$ can be expanded in inverse powers of $P_\mu$ and $M_{\eta_H}$ up to an infinite constant, $\ln(\Delta_{\eta_H}^\infty)$, that will be removed later. Hence, we can apply the covariant expression of the supertrace in~\eqref{eq:covariantSTr} giving
\begin{multline}
\mathrm{STr}\, \ln\Delta_{\eta_H}^{\eminus1}=  \pm \int \dd^dx\int \frac{\dd^dp}{(2\pi)^d}\,\tr\! \bigg\{\int_\infty^1 \dd\xi\,\frac{\dd(\Delta_{\eta_H}^\xi)^{\eminus1}}{\dd\xi}\,\Delta_{\eta_H}^\xi(p_\mu + i\,\tilde{G}_{\mu\nu})-\ln(\Delta_{\eta_H}^\infty)\bigg\} \,.
\end{multline}
Since we are after the hard part of this trace, we can Taylor expand $\Delta_{\eta_H}^\xi(p_\mu + i\,\tilde{G}_{\mu\nu})$ to remove $\tilde{G}_{\mu\nu}$ from the argument. We have
\begin{align}\label{eq:LogTermMaster}
\left.\mathrm{STr}\, \ln\Delta_{\eta_H}^{\eminus1}\right|_\mathrm{hard} &= \pm\int \dd^dx\int \dfrac{\dd^d p}{(2\pi)^d} \tr\! \bigg\{ \int_\infty^1 \dd \xi \frac{\dd(\Delta_{\eta_H}^\xi)^{\eminus1}}{\dd\xi} \,\Delta_{\eta_H}^\xi\sum_{n=1}^{\infty} \big( \mathcal{G}_{\eta_H} \Delta_{\eta_H}^\xi \big)^{\! n}\bigg\} \,,
\end{align}
where we subtracted the $n=0$ term of the series and the infinite constant, which combine to give $\ln\Delta_{\eta_H}^{\eminus1}$ and cancel against the path integral normalization factor. In this expression, we omitted the argument of $\Delta_{\eta_H}^\xi(p_\mu)$ for notational simplicity and defined
\begin{align}\label{eq:Gweird}
\mathcal{G}_i &=\left\{
\begin{array}{ccc}
- i \anticommutator{p^\mu}{\tilde{G}_{\mu\nu}} \partial_p^\nu + (\tilde{G}_{\mu\nu} \,\partial_p^\nu)^2 & \qquad\quad & \mathrm{(scalar)}\\
- i \gamma^{\mu}\, \tilde{G}_{\mu\nu}\, \partial_p^\nu && \mathrm{(fermion)}\\
+ i \anticommutator{p^\mu}{\tilde{G}_{\mu\nu}} \partial_p^\nu - (\tilde{G}_{\mu\nu} \,\partial_p^\nu)^2 & \qquad\quad & \mathrm{(vector)}
\end{array}
\right.\,.
\end{align}
The remaining evaluation of this supertrace is rather arduous but nevertheless straightforward, since the integral in $\xi$ is trivial after performing the well-known loop integrals
	\begin{multline} \label{eq:single_scale_int}
	\mu^{2 \epsilon} \! \int \dfrac{\dd^d p}{(2\pi)^d} \dfrac{p_{\mu_1} \cdots p_{\mu_{2k}}}{ (p^2 -M^2)^{\alpha} p^{2\beta} } \\
	= g_{\mu_1 \ldots \mu_{2k}} \dfrac{(-1)^{\alpha + \beta +k} i}{(4\pi)^2} M^{2(2+k - \alpha - \beta) } \left(\dfrac{\bar{\mu}^2 e^{\gamma_E} }{M^2} \right)^{\!\! \epsilon}  \dfrac{ \Gamma(\tfrac{d}{2} +k - \beta) \Gamma(\alpha +\beta -\tfrac{d}{2} - k)}{ 2^k \Gamma(\alpha) \Gamma( \tfrac{d}{2}+ k) }. 
	\end{multline}
The evaluation of the log-type traces up to dimension six was done e.g. in~\cite{Ball:1988xg}. For completeness, we list them here up to $\mathcal{O}(M_i^{-2})$:
\begin{align}\label{eq:logTraces}
\frac{i}{2}\,\mathrm{STr}\,\ln\,\Delta^{\eminus1}_{\Phi,c_V}\Big|_\mathrm{hard} &= \mp\dfrac{1}{16\pi^2} \tr\! \bigg\{\dfrac{1}{12} \ln\dfrac{\bar{\mu}^2}{M_{\Phi,c_V}^2} \, G^2_{\mu\nu}
+\dfrac{1}{M_{\Phi,c_V}^2} \! \left(\dfrac{1}{60} (D^\mu G_{\mu\nu})^2 + \dfrac{i}{90} G\ud{\mu}{\nu} G\ud{\nu}{\rho} G\ud{\rho}{\mu} \right) \bigg\}, \nonumber \\
\frac{i}{2}\,\mathrm{STr}\,\ln\,\Delta^{\eminus1}_{\Psi}\Big|_\mathrm{hard}& = -\dfrac{1}{16\pi^2} \tr\! \bigg\{\dfrac{1}{3} \ln\dfrac{\bar{\mu}^2}{M_\Psi^2} \, G^2_{\mu\nu}
+ \dfrac{1}{M_\Psi^2} \! \left(\dfrac{2}{15} (D^\mu G_{\mu\nu})^2 - \dfrac{i}{45} G\ud{\mu}{\nu} G\ud{\nu}{\rho} G\ud{\rho}{\mu} \right) \bigg\}, \nonumber \\
\frac{i}{2}\,\mathrm{STr}\,\ln\,\Delta^{\eminus1}_V\Big|_\mathrm{hard} &= \dfrac{1}{16\pi^2} \tr\! \bigg\{ \dfrac{1}{12} \left( 1 - 2 \ln\dfrac{\bar{\mu}^2}{M_V^2} \right)\, G^2_{\mu\nu}
    \nonumber \\
    &\hspace{17ex}- \dfrac{1}{M_V^2} \! \left(\dfrac{1}{30} (D^\mu G_{\mu\nu})^2 + \dfrac{i}{45} G\ud{\mu}{\nu} G\ud{\nu}{\rho} G\ud{\rho}{\mu} \right) \bigg\},
\end{align}
where we removed the divergences in the dimension-four terms using the $\overline{\mathrm{MS}}$ scheme. They can be trivially recovered by taking $\ln \mu^2/M_{\eta_H}^2\to\ln \mu^2/M_{\eta_H}^2+1/\epsilon$ (for $d=4-2\epsilon$).

\subsubsection{Power-type supertraces} \label{subsec:power_traces}
For the power-type traces, it is simpler to show that $Q(P_\mu,U_k)=(\Delta X)^n$ are indeed expansions of rational functions in $P_\mu$ and momentum-independent terms. Indeed, this is clearly the case for $\Delta(P_\mu)$, while in local theories the $X$ interactions can be written as 
\begin{align}\label{eq:Xexpansion}
X(P_\mu,\hat x)=\sum_{n=0}^{\infty}X_n^{\mu_1\cdots\mu_n}(\hat x)\,P_{\mu_1}\cdots P_{\mu_n}\,,
\end{align}
where $X_n$ are functions of fields and derivatives of fields acting inside commutators, such as $[P_\mu,\phi]=i(D_\mu \phi)$. The $P_\mu$ terms in the $X$ expansion are usually termed as ``open covariant derivatives''. Note that the expansion of $X$ as a polynomial in $P_\mu$ is not unique, since terms of the form $[P_\mu,\phi]$ always can be arranged as $[P_\mu,\phi]=P_\mu\,\phi-\phi\,P_\mu$. We fix this ambiguity by arranging the $P_\mu$ operators always to the rightmost.

Having argued that $Q(P_\mu,U_k)=(\Delta X)^n$ are expansions of rational functions in $P_\mu$ and $X_n^{\mu_1\cdots\mu_n}(\hat x)$, we can apply the covariant expression of the supertrace in~\eqref{eq:covariantSTr} giving
\begin{align}
\mathrm{STr}\left[(\Delta X)^n\right]&= \pm\, \int \dd^dx\int \frac{\dd^dp}{(2\pi)^d}\mathrm{tr}\left\{\Big[\Delta(p_\mu + i\,\tilde{G}_{\mu\nu} \,\partial_p^\nu) \,\tilde X\Big]^n\right\}\,,
\end{align}
with $\tilde X\equiv X(p_\mu + i\,\tilde{G}_{\mu\nu} \,\partial_p^\nu,\tilde X_n^{\mu_1\cdots\mu_n}(x))$. Once more, we can benefit from only needing the hard part of the supertrace to expand out the $\tilde{G}_{\mu\nu}$ terms in the propagators, namely,
\begin{align}\label{eq:PowerTermMaster}
\left.\mathrm{STr}\left[(\Delta X)^n\right]\right|_\mathrm{hard}&= \pm\, \int \dd^dx\int \frac{\dd^dp}{(2\pi)^d}\mathrm{tr}\left\{\Big[\Delta\sum_{m=0}^{\infty} \big( \mathcal{G} \Delta \big)^{\! m} \tilde X\Big]^n \right\}\,,
\end{align}
where $\Delta(p_\mu)$ are the free propagators defined in~\eqref{eq:Prop}, and $\mathcal{G}$ is defined in~\ref{eq:Gweird}. As with the log-type supertrace, the remaining evaluation of the power-type supertrace is straightforward and the loop integrals can be readily evaluated with 
    \begin{multline}\label{eq:MasterIntegral}
    \int \dfrac{\dd^d p}{(2\pi^d)} \dfrac{p_{\mu_1} \cdots p_{\mu_{2k}}}{(p^2 -M^2_1)^{\alpha_1} \cdots (p^2 -M^2_n)^{\alpha_n} p^{2\beta} } \\
    = \sum_{m=1}^n \sum_{k=0}^{\alpha_m -1} \dfrac{1}{k!} \int \dfrac{\dd^d p}{(2\pi^d)} \dfrac{p_{\mu_1} \cdots p_{\mu_{2k}}}{ (p^2 -M^2_m)^{\alpha_m-k} p^{2\beta} } \left( \dfrac{\partial}{\partial M_m^2} \right)^{\!\!k} \prod_{\ell \neq m} \dfrac{1}{ (M_m^2 - M_\ell^2)^{\alpha_\ell} }\,,
    \end{multline}
along with formula~\eqref{eq:single_scale_int}. However, the amount of algebra involved in evaluating these supertraces makes it rather tedious without the use of computer tools.

\subsection{Comments on the treatment of \texorpdfstring{$\gamma_5$}{gamma5} in fermion supertraces}

There is an added complication in the evaluation of fermion supertraces in terms where all propagators are fermionic, resulting in traces of $\gamma$-matrices. One of the primary outstanding problems in dimensional regularization is how to continue the definition of $ \gamma_5 $ away from 4 dimensions, cf.~\cite{Jegerlehner:2000dz}. Whatever regularization procedure (and renormalization scheme) is used in the matching calculation requires the same choice to be used in subsequent computations in the EFT. We therefore propose to use a semi-naive implementation of dimensional regularization, as Naive Dimensional Regularization (NDR) often is the most practical choice for perturbative calculations. 

For the Dirac algebra we formally set 
    \begin{equation}
    \anticommutator{\gamma^\mu}{\gamma^\nu} = 2 g^{\mu \nu}, \qquad \anticommutator{\gamma^\mu}{\gamma_5} = 0, \andeq \gamma_5^2 = \mathbb{1}, 
    \end{equation}
where all Lorentz indices are $ d $-dimensional. This algebra in conjunction with cyclicity of the trace results in the vanishing of all traces with an odd number of $ \gamma_5 $'s, prohibiting the recovery of the four-dimensional result in the limit $ d \to 4 $ . We therefore abandon the cyclicity of \go traces, while formally substituting~\cite{Chanowitz:1979zu,Mihaila:2012pz}
    \begin{equation} \label{eq:SNDR_pescription}
    \tr[ \gamma^\mu \gamma^\nu \gamma^\rho \gamma^\sigma \gamma_5 ] = - 4 i \varepsilon^{\mu \nu \rho \sigma} + \ord{\epsilon}. 
\end{equation}
With this prescription the choice of where the \go traces are read from---meaning which $\gamma$ in a Dirac trace is written as the left-most---results in another $\ord{\epsilon} $ ambiguity. This ambiguity is therefore only manifest in divergent diagrams, where the $ \epsilon $ pole and the $ \ord{\epsilon} $ trace ambiguity combines to give a finite ambiguity in the computation. 

The calculation of the one-loop effective action in the UV theory does not involve any UV divergent \go diagram due to anomaly cancellation and the prescription is unambiguous. A complication arises when performing the matching computation and identifying $ \L^{\scriptscriptstyle (1)}_\mathrm{EFT} $ with the hard part of the functional supertrace: the expansion of the loop integral in heavy masses and hard loop momenta can introduce spurious IR divergences in some of the integrals. The IR divergences combined with the reading-point ambiguity introduce an ambiguity in $\L^{\scriptscriptstyle (1)}_\mathrm{EFT} $. Conveniently, when expanding loop integrals by regions, spurious IR divergences in the hard part of the integral are known to cancel exactly against corresponding UV divergences in the soft part, which in our case corresponds to 1-loop diagrams in the EFT. With the $ \epsilon $ poles canceling in $ \Gamma^{\scriptscriptstyle (1)}_\EFT $, so too will the ambiguities in \go diagrams, as long as the $ \gamma_5 $ prescription is applied consistently between the matching and the EFT calculations. That is, as long as the reading points are chosen identically.

The consistent choice of reading point is perhaps best illustrated with an example. If the UV theory involves both light and heavy fermions, $\psi$ and $\Psi$, the effective action can contain contributions of the form 
    \begin{equation}
    \Gamma^{\scriptscriptstyle (1)}_\mathrm{UV} \supset -\frac{i}{2}\,\mathrm{STr} \big[ \Delta_\psi\,X_{\psi\Psi}\, \Delta_{\Psi}\,X_{\Psi\psi} \big]. 
    \end{equation}
In the UV theory the \go piece of this part of the effective action is finite, thereby ensuring that there is no ambiguity from the reading point of the Dirac trace, which is embedded in the supertrace. Computing equivalent one-loop amplitudes in the EFT will involve a part coming from one-loop contributions to $ S_\mathrm{EFT} $ and one loop diagrams with the tree-level EFT:
    \begin{equation}
    \Gamma^{\scriptscriptstyle (1)}_\mathrm{EFT} \supset S_\mathrm{EFT}^{\scriptscriptstyle (1)} -\frac{i}{2}\, \mathrm{STr} \big[ \Delta_\psi\,X^{\mathrm{EFT}}_{\psi\psi} \big],
    \end{equation}
where the corresponding contribution to the EFT fluctuation operator quickly is identified as 
    \begin{equation}
    X^{\mathrm{EFT}}_{\psi\psi} \supset X_{\psi\Psi}\, \frac{i \slashed D + M_\Psi}{M_\Psi^2} \,X_{\Psi\psi} +\ldots 
    \end{equation}
The two contributions to $ \Gamma^{\scriptscriptstyle (1)}_\mathrm{EFT}$ are readily identified with the hard and soft part of the UV loops, respectively. Consequently, the $ \epsilon $ poles cancel between them and, when the same reading point is chosen, so will the reading point ambiguity in the finite part of the effective action. The reading point can be fixed by e.g. making sure that $ X_{\Psi\psi} $ is the last piece of the trace, in both EFT computation and matching computations.

\subsection{Summary of the functional matching method}
\label{subsec:summaryPI}
In this section, we summarize the relevant steps needed to perform functional EFT matching at the one-loop level. These are:
\begin{itemize}
    \item[i)] \textbf{Collecting all fields, heavy and light, into field multiplets:} To obtain the fluctuation operator~\eqref{eq:flucOp}, one needs to take functional derivatives with respect to all fields in the theory, including field conjugates in the case of complex fields. For this reason, it is useful to arrange the fields into field multiplets
    \begin{align}\label{eq:ComplexForm}
    \varphi_\phi&=
    \begin{pmatrix}
    \phi\\
    \phi^*
    \end{pmatrix}
    \,,&
    \varphi_\psi&=
    \begin{pmatrix}
    \psi\\
    \psi^c
    \end{pmatrix}
    \,,&
    \varphi_A&=
    \begin{pmatrix}
    A_\mu\\
    A_\mu^*
    \end{pmatrix}
    \,,
    \end{align}
    in the case of complex scalars, fermions, and complex vectors, respectively. Here $f^c=C \bar{f}^\intercal$ is the charge-conjugated fermion with $C$ being the charge conjugation matrix and both $f$ and $f^c$ 4-component Dirac spinors. In the case when only some chiralities are present, like in the SM, chiral projectors should be used in the corresponding interactions. Furthermore, it is convenient to organize the fields (in the form of~\eqref{eq:ComplexForm}) into one field multiplet for each field type. These types are heavy scalar, light scalar, heavy fermion, light fermion, heavy vector, light vector, heavy ghost, and light ghost, which we generically denote by $\Phi$, $\phi$, $\Psi$, $\psi$, $V$, $A$, $c_V$, and $c_A$, respectively.
    
    \item[ii)] \textbf{Obtaining the heavy field EOMs:} The EOMs can be determined directly by setting the first functional derivative of the UV Lagrangian with respect to the heavy fields equal to zero, namely
    \begin{align}
    \left.\frac{\delta\L_\UV}{\delta\eta_H}\right|_{\eta=\hat\eta}=0\,,
    \end{align}
    where we remind the reader that the hat denotes field configurations that satisfy the EOMs, and $\eta_H$ contains all the heavy fields multiplets $\eta_\Phi$, $\eta_\Psi$, $\eta_V$, and/or $\eta_{c_V}$ of the theory. These equations need to be expanded to a given order in the heavy mass expansion, matching the desired order in the EFT expansion, to obtain order-by-order expressions of the heavy fields in terms of light fields. The tree-level EFT Lagrangian is obtained by replacing these expressions into the UV Lagrangian.

    \item[iii)] \textbf{Determining the \textit{X} terms:} These are obtained from the second functional derivative of the UV Lagrangian with respect to heavy and light fields after subtraction of the inverse propagators. More precisely,
    \begin{align}\label{eq:XtermExpr}
    X_{ij}=\delta_{ij}\,  \Delta_i^{\eminus1}-\left.\frac{\delta^2\L_\UV}{\delta\bar\eta_i\,\delta\eta_j}\right|_{\eta=\hat\eta}\,,
    \end{align}
    with the inverse propagators given in~\eqref{eq:Prop}. Light-particle masses are always included in the $X$ terms to better organize the power counting. If the $X$ terms contain derivative interactions, these should be arranged in the form of~\eqref{eq:Xexpansion}. In most practical cases, only the terms $X_0$ and/or $X_1^\mu$ of this expansion are present. It is convenient to keep track of the mass dimension of the fields and derivatives acting on fields (e.g. ``close covariant derivatives'') inside each $X$ term, since this provides a simple power counting for the EFT expansion.
    
    \item[iv)] \textbf{Identifying and evaluating the relevant supertraces:} The next step is to identify the relevant log-type and power-type supertraces that enter into the one-loop EFT matching equation~\eqref{eq:Master}. Log-type supertraces are model-independent, since they do not depend on the $X$ terms (encoding the relevant $\L_\UV$ interactions), but only on the heavy-field propagators. They can be evaluated from the expansion~\eqref{eq:LogTermMaster}. A list of all log-type supertraces evaluated up to operators of dimension six is given in~\eqref{eq:logTraces}. A log-type supertrace should be included for each of the heavy fields in the theory, including the complex conjugate in the case of complex fields.
    
    Power-type supertraces do depend on the $X$ terms and should (a priori) be computed for every UV model. The infinite series in~\eqref{eq:Master}, and hence the number of supertraces to compute, is truncated by the desired mass dimension of the EFT operators, which is determined by adding the mass dimensions of each of the $X$ terms appearing in a given supertrace. An important subtlety to consider is that the series~\eqref{eq:Master} gives rise to symmetry factors in some of the supertraces. These symmetry factor are given by the inverse of the number of times the trace repeats itself under cyclic permutations. These types of supertraces can be evaluated by means of the expansion in~\eqref{eq:PowerTermMaster}. 
\end{itemize}

The procedure presented here closely follows the prescription presented in~\cite{Cohen:2020fcu}. However, there are a number of differences between the two. First, light-field masses are included in the $X$ terms and not in the propagators. Second, the derivative expansion of $X$ is defined such that \textit{all} derivatives are made to act to the rightmost, c.f.~\eqref{eq:Xexpansion}. Furthermore, we do not adopt a diagrammatic description for the identification of the relevant power-type supertraces, since this task is performed automatically by \pkg{SuperTracer}, as we describe below.

\section{SuperTracer in a nutshell}
\label{sec:SuperTracer}

\begin{figure}[t]
    \centering
    \includegraphics[width=\textwidth]{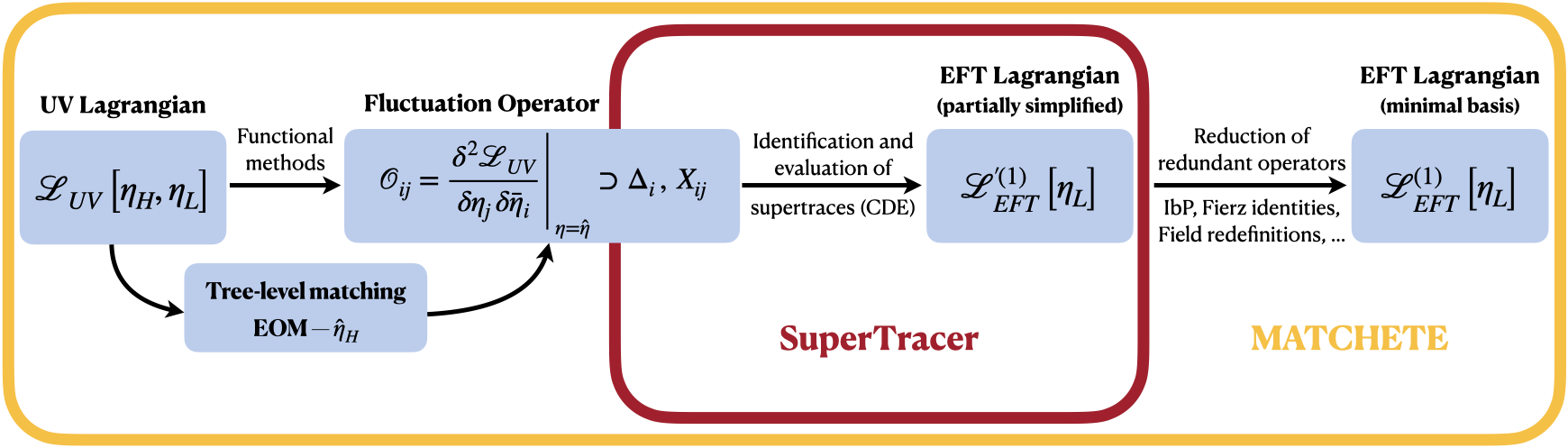}
    \caption{Summary of the functional approach to one-loop EFT matching. Highlighted in red, we show the parts of the procedure that are taken care of by \pkg{SuperTracer}, and in yellow those parts that will be handled by the master program \pkg{MATCHETE} (to be released).}
    \label{fig:matchingflow}
\end{figure}

\pkg{SuperTracer} is a \pkg{Mathematica} package aimed at facilitating the functional EFT matching procedure described in Section~\ref{sec:method} and illustrated in Figure~\ref{fig:matchingflow}. The package takes over the most tedious parts of this procedure by automating the process of identifying and evaluating all relevant supertraces. The program also partially simplifies the resulting operators, as described below. However, it does not provide a complete reduction to a basis, and the calculation of the $X$ interactions and heavy field EOMs still has to be done manually. We delegate these tasks to the \pkg{Mathematica} package \pkg{MATCHETE} (\textbf{Match}ing \textbf{E}ffective \textbf{T}heories \textbf{E}fficiently)~\cite{MATCHETE}, which we are currently developing, and which will include \pkg{SuperTracer} at its core. The ultimate goal is to fully automate the matching procedure, having as input a user-defined UV Lagrangian, and completely eliminating the need for manually determining and inserting the $X$ interactions.

The main routines in the current implementation of \pkg{SuperTracer} evaluate log- and power-type supertraces by performing the following steps:
\begin{enumerate}[i)]
    \item The propagators are reconstructed from the input list of $X$ interactions, and everything is placed in a non-commutative product. Fermionic traces are assigned an extra factor of $(-1)$.
    \item The covariant expansion of $\Delta$ and $X$ terms are performed to the appropriate order. All momentum derivatives act through the expression to terminate on the right.
    \item All Dirac products are simplified and matched to a basis of anti-symmetrized products, $ \Gamma_{\mu_1,\ldots \mu_n} = \gamma_{[\mu_1}\cdots \gamma_{\mu_n]} $, and the loop integrals are evaluated using dimensional regularization with $d=4-2\epsilon$ in the $ \overline{\mathrm{MS}} $ scheme.
\end{enumerate}
Added utility is provided by allowing the user to substitute model-specific expressions in the $X$ operators, making it possible to directly perform additional simplifications such as evaluating Dirac traces. 

As for validation of the package, we have cross-checked a variety of supertraces against the \pkg{STrEAM} package as kindly provided by the authors~\cite{Cohen:2020qvb}. Furthermore, the two example models discussed in Section~\ref{sec:examples} have allowed us to check the package against a sample diagrammatic computation (cf. Appendix~\ref{app:VLexample}) and previous literature.  

\subsection{Downloading and installing the package}

The \pkg{SuperTracer} package is free software under the terms of the GNU General Public License v3.0 and is publicly available in the GitLab repository
\begin{center}
\href{https://gitlab.com/supertracer/supertracer}{https://gitlab.com/supertracer/supertracer}
\end{center}
The package can be installed in one of two ways:
\begin{enumerate}[i)]
    \item \textit{Automatic installation}: The simplest way to download and install \pkg{SuperTracer} is to run the following command in a \pkg{Mathematica} notebook:
\begin{mmaCell}{Input}
  Import["https://gitlab.com/supertracer/supertracer/-/raw/
  master/install.m"]
\end{mmaCell}
    This will download and install \pkg{SuperTracer} in the \textit{Applications} folder of \pkg{Mathematica}'s base directory.
    
    \item \textit{Manual installation}: The user can also manually download the package from the GitLab repository \href{https://gitlab.com/supertracer/supertracer}{\color{red}\faicon{gitlab}}. We recommend placing the \pkg{SuperTracer} folder in the \textit{Applications} folder of \pkg{Mathematica}'s base directory, so its location does not need be specified before loading the package. Alternatively, the user can place the \pkg{SuperTracer} folder in a different directory and specify its location via
\begin{mmaCell}{Input}
  AppendTo[\$Path,"directory"];
\end{mmaCell}
    with \texttt{\color{DarkGray}directory} being the path to the \pkg{SuperTracer} folder.
    
\end{enumerate}
Once installed, the user can load \pkg{SuperTracer} in any \pkg{Mathematica} notebook by running
\begin{mmaCell}{Input}
  << \mmaDef{SuperTracer\(\,\grave\,\)}
\end{mmaCell}

\vspace{-5pt}
\includegraphics[width=0.96\textwidth]{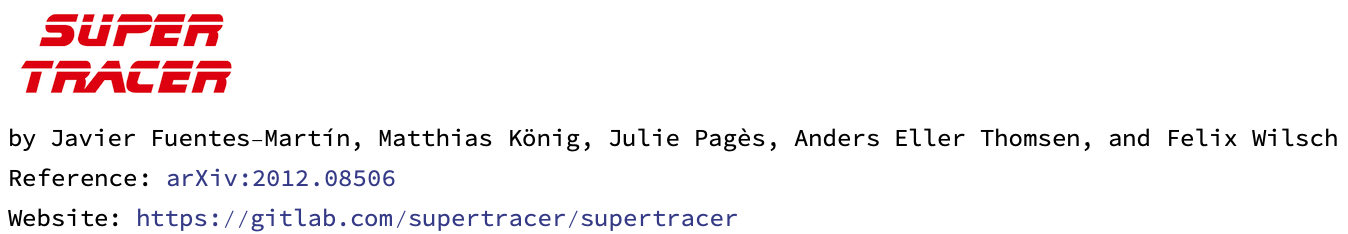}

\begin{table}[]
    \centering
    \renewcommand{\arraystretch}{1.2}
    \begin{tabular}{p{.25\textwidth} p{.69\textwidth}}
    \textbf{Variable} & \textbf{Description}  \\
    \toprule
    \mmaInlineCell[]{Input}{\(\Phi,\phi,\Psi,\psi\),\mmaDef{V},\mmaDef{A},\mmaDef{cV},\mmaDef{cA}} & Field types. They are, respectively, heavy scalar, light scalar, heavy fermion, light fermion, heavy vector, light vector, heavy ghost, and light ghost.~\medskip\\
    
    \mmaInlineCell[]{Input}{\mmaDef{X}[\{f1,f2\},\mmaUnd{<ord>}]} & Input form for the $X$ interactions. The arguments \mmaInlineCell[]{Input}{f1} and \mmaInlineCell[]{Input}{f2} should be field types, while \mmaInlineCell[]{Input}{\mmaUnd{<ord>}} is an optional argument specifying the order of $X$, which can be a single number (if $X_0$ is the only term in the expansion in~\eqref{eq:Xexpansion}) or a list of numbers indicating the orders of the $X_{0,1,2,\dots}$ terms. If no \mmaInlineCell[]{Input}{\mmaUnd{<ord>}} is given, the default values in \mmaInlineCell[]{Input}{\mmaDef{Xords}} are taken. This variable has a special output format, e.g. \mmaInlineCell[]{Input}{\mmaDef{X}[\{\(\psi\),\(\psi\)\}]} shows as \mmaInlineCell[]{Input}{\mmaSub{\mmaDef{X}}{\(\psi\)\(\psi\)}} and \mmaInlineCell[]{Output}{X[\{\(\psi\),\(\psi\)\},2]} as \mmaInlineCell[]{Output}{\mmaSubSup{X}{\(\psi\)\(\psi\)}{[2]}}.~\medskip\\
    
    \mmaInlineCell[]{Input}{\mmaDef{Xords}} & Association with the default interaction order of each $X$ term, e.g. \mmaInlineCell[]{Input}{\mmaDef{Xords}@\{\(\psi\),\(\psi\)\}} returns \mmaInlineCell[]{Output}{1}.~\medskip\\
        
    \mmaInlineCell[]{Input}{\mmaDef{STr}[Xterms]} & A header denoting a supertrace of the list \mmaInlineCell[]{Input}{Xterms} of $X$ interactions. Symmetry factors and a global factor of $-\frac{i}{2}$ is included into the definition of this variable for notational simplicity.~\medskip\\
    
    \mmaInlineCell[]{Input}{\mmaDef{M}[label]} & Heavy field mass. The output has special format \mmaInlineCell[]{Output}{\mmaSub{M}{label}} \medskip\\
        
    \mmaInlineCell[]{Input}{\mmaDef{\$DegenerateMasses}} & Global variable that can be \mmaInlineCell[]{Input}{True} or \mmaInlineCell[]{Input}{False} (set to \mmaInlineCell[]{Input}{True} by default). If \mmaInlineCell[]{Input}{True}, all heavy scales are assumed to be equal to \mmaInlineCell[]{Input}{\mmaDef{M}["H"]} (output format \mmaInlineCell[]{Output}{\mmaSub{M}{H}}).\medskip\\
    
    \mmaInlineCell[]{Input}{\mmaDef{\(\mu\)bar2}} & Matching scale squared. This variable has the special output format \mmaInlineCell[]{Output}{\mmaSup{\(\overline{\mu}\)}{2}}.\medskip\\
    
    \mmaInlineCell[]{Input}{\mmaDef{LF}[\{\mmaUnd{\mmaSub{m}{1}},...\},\{\mmaUnd{\mmaSub{n}{1}},...\}]} & Finite part of the loop integral with propagators $(p-m_1)^{-2n_1}\dots(p-m_k)^{-2n_k}p^{-2n_{k+1}}$, where $n_1,\dots,n_{k+1}$ are integer numbers. This variable has the special output format \mmaInlineCell[]{Output}{\mmaSub{LF}{\mmaSub{n}{1},...}[\mmaSub{m}{1},...]}. \\
    \bottomrule
    \end{tabular}
    \caption{Main \pkg{SuperTracer} variables.}
    \label{tab:MainVariables}
\end{table}

\subsection{SuperTracer global variables and routines}

After the package has been loaded, a variety of global variables and routines are defined. The main global variables are described in Tables~\ref{tab:MainVariables} and~\ref{tab:XVariables}. The routines available to the user are:

\newpage
\medskip\noindent
\textbf{Main SuperTracer routines}
\begin{itemize}
    \item \rut{LogTerm[field,<order>]} returns the log-type terms, resulting from integrating out a heavy \textit{field}, to a given \textit{order} in the EFT expansion. The \textit{field}, which can be  \mmaInlineCell[]{Input}{\mmaDef{\(\Phi\)},\mmaDef{\(\Psi\)},\mmaDef{V}} or \mmaInlineCell[]{Input}{\mmaDef{cV}}, is assumed to be real (or Majorana in the fermionic case), so the output should be multiplied by two in the case of complex (or Dirac) fields. The \textit{order} is assumed to be inclusive unless it is given inside curly brackets, e.g. \mmaInlineCell[]{Input}{\mmaDef{LogTerm}[\(\Phi\),6]} provides all operators up to dimension six, while \mmaInlineCell[]{Input}{\mmaDef{LogTerm}[\(\Phi\),\{6\}]} provides only operators of dimension six. The \textit{order} is an optional argument. If no \textit{order} is given, \mmaInlineCell[]{Input}{6} is assumed. 
    
    \item \rut{PowerTerms[Xterms,<order>]} returns the sum of all power-type traces that need to be computed for a given list of $X$ \textit{terms} to a given \textit{order} in the EFT expansion. As for \rut{LogTerm}, the \textit{order} is an optional argument and it is treated in the same manner. For example, if we have a theory with three $X$ terms, $X_{\psi\Phi}^{\scriptscriptstyle[3/2]}$, $X_{\Phi\psi}^{\scriptscriptstyle[3/2]}$ and $X_{\psi\psi}^{\scriptscriptstyle[3]}$, by running \rut{PowerTerms}
    \begin{mmaCell}{Input}
    \mmaDef{PowerTerms}[\{\mmaDef{X}[\{\(\psi\),\(\Phi\)\},3/2],\mmaDef{X}[\{\(\psi\),\(\psi\)\},3]\}]
    \end{mmaCell}
    \begin{mmaCell}{Output}
    \mmaDef{STr}[\{\mmaSubSup{X}{\(\Phi\)\(\psi\)}{[3/2]},\mmaSubSup{X}{\(\psi\)\(\Phi\)}{[3/2]}\}] + \mmaDef{STr}[\{\mmaSubSup{X}{\(\Phi\)\(\psi\)}{[3/2]},\mmaSubSup{X}{\(\psi\)\(\psi\)}{[3]},\mmaSubSup{X}{\(\Psi\)\(\Phi\)}{[3/2]}\}] +
    \mmaDef{STr}[\{\mmaSubSup{X}{\(\Phi\)\(\psi\)}{[3/2]},\mmaSubSup{X}{\(\psi\)\(\Phi\)}{[3/2]},\mmaSubSup{X}{\(\Phi\)\(\psi\)}{[3/2]},\mmaSubSup{X}{\(\psi\)\(\Phi\)}{[3/2]}\}]
    \end{mmaCell}
    we find that three supertraces, denoted by \rut{STr}, need to be computed at the level of dimension six operators. Note that the symmetry factor $1/2$ that would appear in \mmaInlineCell[]{Output}{\mmaDef{STr}[\{\mmaSubSup{X}{\(\Phi\)\(\psi\)}{[3/2]},\mmaSubSup{X}{\(\psi\)\(\Phi\)}{[3/2]},\mmaSubSup{X}{\(\Phi\)\(\psi\)}{[3/2]},\mmaSubSup{X}{\(\psi\)\(\Phi\)}{[3/2]}\}]} (cf. Section~\ref{subsec:summaryPI}) and a global $-\frac{i}{2}$ is absorbed into the definition of \rut{STr} for notational simplicity. Further note that conjugate interactions need not be introduced since these are automatically included by \rut{PowerTerms}. Indeed, in our example we have input \mmaInlineCell[]{Input}{\mmaDef{X}[\{\(\psi\),\(\Phi\)\},3/2]} but not \mmaInlineCell[]{Input}{\mmaDef{X}[\{\(\Phi\),\(\psi\)\},3/2]}.
    
    \item \rut{STrTerm[Xterms,<order>,<Xsubstitutions>]} evaluates the power-type supertrace of a given list of $X$ \textit{terms} to a given \textit{order} in the EFT expansion. The output of \rut{STrTerm} is assumed to be inside $\int\,\dd^dx\;\frac{1}{16\pi^2}\tr\!{[.]}$, with $\tr\!{[.]}$ being a trace over internal degrees of freedom. Moreover, note that the definition of supertrace in \pkg{SuperTracer} includes symmetry factors and a global $-\frac{i}{2}$ factor. This routine further allows for the optional substitution of the $X$ \textit{terms} into their explicit expressions in terms of fields. Rather than explaining how to perform $X$ \textit{substitutions} here, we provide detailed usage examples of this functionality in Section~\ref{sec:examples} and in the ancillary \pkg{Mathematica} notebooks. The \textit{order} is treated in the same way as in the \rut{LogTerm} routine. However, if the list of $X$ \textit{substitutions} is given, the \textit{order} must also be given. Finally, the $\epsilon$ poles are removed from the output of \rut{STrTerm}. They can be easily recovered since their coefficient matches that of the renormalization-scale logarithm. To simplify the evaluation of (especially) multi-scale integrals, the finite part of the loop functions are kept implicit under the variable \pkg{LF}. \rut{EvaluateLoopFunctions[expr]} evaluates all loop functions in the expression \mmaInlineCell[]{Input}{expr}.
    
    \item \rut{SuperSimplify[expr]} is the primary simplification routine, which provides a one-point-stop for simplifications of \pkg{SuperTracer} outputs. It
    simplifies outputs of \rut{STrTerm} and \rut{LogTerm} by attempting different index labels and collecting terms with the same operator structure. \rut{SuperSimplify} also calls \rut{SimplifyOutput}, which uses integration by parts, commutator, and Jacobi identities to match the output to a basis of operators.  
    Although the outputs of \pkg{SuperTracer} is shown as a normal sum of terms to the user, its full \pkg{Mathematica} form consists of a sum of \mmaInlineCell[]{Input}{\mmaDef{LTerm}[coeff,OpStr]} that separate the coefficient and operator structure of each term for better internal manipulations.
\end{itemize}

\begin{table}[]
    \centering
    \renewcommand{\arraystretch}{1.2}
    \begin{tabular}{p{.35\textwidth} p{.58\textwidth}}
    \textbf{Variable} & \textbf{Description}  \\
    \toprule
     
    \mmaInlineCell[]{Input}{\mmaDef{g}[\mmaUnd{\(\mu\)},\mmaUnd{\(\nu\)}]}& Lorentz metric tensor. The arguments \mmaInlineCell[]{Input}{\mmaUnd{\(\mu\)},\mmaUnd{\(\nu\)}} are Lorentz indices. This variable has the special output format~\mmaInlineCell[]{Output}{\mmaSub{\mmaDef{g}}{\(\mu\nu\)}}.~\medskip\\
     
    \mmaInlineCell[]{Input}{\(\varepsilon\)[\mmaUnd{\(\mu\)},\mmaUnd{\(\nu\)},\mmaUnd{\(\rho\)},\mmaUnd{\(\sigma\)}]}& Levi-Civita symbol. The arguments \mmaInlineCell[]{Input}{\mmaUnd{\(\mu\)},\mmaUnd{\(\nu\)},\mmaUnd{\(\rho\)},\mmaUnd{\(\sigma\)}} are Lorentz indices. This variable has the special output format \mmaInlineCell[]{Output}{\mmaSub{\(\varepsilon\)}{\(\mu\nu\rho\sigma\)}}.~\medskip\\
    
    \mmaInlineCell[]{Input}{\(\gamma\)[\mmaUnd{\(\mu\)}], \mmaUnd{\(\gamma\)}[5]} & Dirac matrices. The argument \mmaInlineCell[]{Input}{\mmaUnd{\(\mu\)}} is a Lorentz index. The output has the special form \mmaInlineCell[]{Output}{\mmaSub{\(\gamma\)}{\(\mu\)}} or \mmaInlineCell[]{Output}{\mmaSub{\(\gamma\)}{5}}.~\medskip\\
    
    \mmaInlineCell[]{Input}{\mmaDef{PL}, \mmaDef{PR}} & Chiral projectors. The output has the special output form \mmaInlineCell[]{Output}{\mmaSub{P}{L}} and \mmaInlineCell[]{Output}{\mmaSub{P}{R}}, respectively. \medskip\\

     \mmaInlineCell[]{Input}{\mmaDef{T}[\{repA[A],rep[a],rep[b]\}]} & Symmetry generator. The arguments \mmaInlineCell[]{Input}{a} and \mmaInlineCell[]{Input}{b} are indices in the representation \mmaInlineCell[]{Input}{rep}, whereas \mmaInlineCell[]{Input}{A} is in the representation \mmaInlineCell[]{Input}{repA}. The output has the special form~\mmaInlineCell[]{Output}{\mmaSup{T}{Aab}}.~\medskip\\

    \mmaInlineCell[]{Input}{\mmaDef{GA}[\{\mmaUnd{\(\mu\)},\mmaUnd{\(\nu\)}\},\mmaUnd{tag}]} & Field-strength tensor for Abelian groups. The arguments \mmaInlineCell[]{Input}{\mmaUnd{\(\mu\)}} and \mmaInlineCell[]{Input}{\mmaUnd{\(\nu\)}} are Lorentz indices and \mmaInlineCell[]{Input}{\mmaUnd{tag}} is the symbol used to label the Abelian symmetry. Abelian field-strength tensors are displayed as \mmaInlineCell[]{Output}{\mmaSubSup{F}{tag}{   \(\mu\)\(\nu\)}} in the output.~\medskip\\
    
     \mmaInlineCell[]{Input}{\mmaDef{GnA}[\{\mmaUnd{\(\mu\)},\mmaUnd{\(\nu\)}\},\{rep[a],rep[b]\}]} & Field-strength tensor for non-Abelian groups. As before, the arguments \mmaInlineCell[]{Input}{\mmaUnd{\(\mu\)}} and \mmaInlineCell[]{Input}{\mmaUnd{\(\nu\)}} are Lorentz indices, while \mmaInlineCell[]{Input}{a} and \mmaInlineCell[]{Input}{b} label non-Abelian group indices in representation \mmaInlineCell[]{Input}{rep}. Non-Abelian field-strength tensors have output format \mmaInlineCell[]{Output}{\mmaSubSup{G}{rep}{ab \(\mu\)\(\nu\)}}.~\medskip\\
    
    \mmaInlineCell[]{Input}{\(\delta\)[\{rep[a],rep[b]\}]} & Kronecker delta. The two arguments \mmaInlineCell[]{Input}{a} and \mmaInlineCell[]{Input}{b} are indices of the representation \mmaInlineCell[]{Input}{rep}. This variable has the special output format \mmaInlineCell[]{Output}{\mmaSup{\(\delta\)}{ab}}.~\medskip\\
    
    \mmaInlineCell[]{Input}{\mmaDef{eps}[\{rep[a],rep[b]\}]} & Anti-symmetric tensor with two indices. The two arguments \mmaInlineCell[]{Input}{a} and \mmaInlineCell[]{Input}{b} are indices of the representation \mmaInlineCell[]{Input}{rep}. This variable has the special output format~\mmaInlineCell[]{Output}{\mmaSubSup{\(\epsilon\)}{rep}{ab}}.~\medskip\\
    
    \mmaInlineCell[]{Input}{\mmaDef{Flavor}[index]}& Flavor index. The header \mmaInlineCell[]{Input}{\mmaDef{Flavor}} labels the \textit{flavor} representation and specifies that \mmaInlineCell[]{Input}{index} is a flavor index. \\

    \bottomrule
    \end{tabular}
    \caption{\pkg{SuperTracer} variables relevant for $X$ substitutions.}
    \label{tab:XVariables}
\end{table}

\medskip\noindent
\textbf{Routines for $\boldsymbol{X}$ substitutions}
\begin{itemize}
    \item \rut{AddField[label,type,<charge(s)>,<countingDim>]} defines a field of a given \textit{type} (cf. Tab.~\ref{tab:MainVariables}) with a given \textit{label}, so it can be used in an $X$ \textit{substitution}. If the field is charged under a single gauge $\U(1)$, its \textit{charge} should be provided as \mmaInlineCell[]{Input}{label[charge]}, where \mmaInlineCell[]{Input}{label} is a label for the $\U(1)$ symmetry chosen by the user and \mmaInlineCell[]{Input}{charge} is a number specifying the field charge. On the other hand, if the field is charged under multiple $\U(1)$ gauge groups, the user should give a list of \textit{charges} with the format \mmaInlineCell[]{Input}{\{label1[charge1],label2[charge2],...\}}. As an example, let us define a heavy scalar field $f$ with charge $2$ under a gauge $\U(1)_L$ that we label by $L$:
    \begin{mmaCell}{Input}
    \mmaDef{AddField}[f,\mmaDef{\(\Phi\)},L[2]]
    \end{mmaCell}
    This creates the field routine \rut{f[Indices]} where the flavor and gauge \textit{indices} carried by the field should be given as a list. If the field carries no indices, no argument or an empty list can be given. In the case of a vector field, the first entry in the list must be a Lorentz index~\mmaInlineCell[]{Input}{\mmaUnd{\(\mu\)}}, i.e. the \textit{indices} should then be given in the format \mmaInlineCell[]{Input}{\{\mmaUnd{\(\mu\)},rep1[ind1],rep2[ind2],...\}}, where \mmaInlineCell[]{Input}{rep1} denotes the representation of the index \mmaInlineCell[]{Input}{ind1}.
    To remove the field \mmaInlineCell[]{Input}{\mmaDef{f}} from the set of defined fields, the routine \rut{RemoveField[f]} can be used.
 
    \item \rut{Bar[obj]} returns the bar of a fermion field or the complex conjugate of other fields. Applying \rut{Bar} to couplings and generators yields their conjugate. The routine can also be applied to representations and charges in $X$ \textit{substitutions} to denote their conjugate. This routine has a special output format, i.e. \mmaInlineCell[]{Input}{\mmaDef{Bar}[obj]} shows as \mmaInlineCell[]{Output}{\(\overline{obj}\)}.
    
    \item \rut{Transp[obj]} returns the transpose of any \textit{object} in Dirac space, that is, fermion fields, chiral projectors, or Dirac matrices. This routine has a special output format, i.e. \mmaInlineCell[]{Input}{\mmaDef{Transp}[obj]} shows as \mmaInlineCell[]{Output}{\mmaSup{obj}{\mmaDef{T}}}. 

    \item \rut{CConj[field]} returns the charge conjugate of a fermion \textit{field}, e.g.  \mmaInlineCell[]{Input}{\mmaDef{CConj}[\mmaDef{f}[]]} gives \mmaInlineCell[]{Input}{\mmaDef{CC}**\mmaDef{\mmaSup{\(\overline{\mathtt{f}}\)}{T}}} with \mmaInlineCell[]{Input}{\mmaDef{CC}} being the charge conjugation matrix. 
    If this routine is applied to something other than a fermion field, the output is aborted and a warning is issued.

    \item \rut{CD[index,expr]} or \rut{CD[\{indices\},expr]} returns the covariant derivative(s) of a given \textit{expression}, with the number of Lorentz \textit{indices} determining the number of derivatives. If the covariant derivative acts on a undefined variable, it is assumed to be vanishing.
\end{itemize}
To keep track of non-commutative objects \pkg{SuperTracer} co-opts \pkg{Mathematica}'s build in \mmaInlineCell[]{Input}{NonCommutativeMultiply}~(\mmaInlineCell[]{Input}{**}). Field objects, elements of the Dirac algebra and field-strength tensors are treated as non-commutative until the end of the computation. Only when using the substitution capability of \rut{STrTerm} is non-commutativity for bosonic fields and field-strength tensors relaxed. All substitution rules must be given as non-commutative products. 

As we have already described in certain routines and global variables, we have defined special output formats for some expressions to facilitate the reading of \pkg{SuperTracer} outputs. The explicit \pkg{Mathematica} expression of the output can be seen by applying the \pkg{InputForm}/\pkg{FullForm} routine. Although understanding this explicit form is not necessary to use all \pkg{SuperTracer} features, it is required when doing further manipulations of the output. We refer the interested reader to Appendix~\ref{app:GlobalPro} for more details on the variables that are used there.

\section{Usage examples}
\label{sec:examples}

Here we illustrate the matching procedure described in Section~\ref{sec:method} and the functionality of the \pkg{SuperTracer} package with two examples of heavy field integration: a toy model with a heavy vector-like fermion and an $S_1$ scalar leptoquark extension of the SM.

\subsection{Toy model with a heavy vector-like fermion}
\label{sec:VLexmaple}

As a first example, we consider a toy model with a heavy fermion $\Psi$ charged under a gauged $\U(1)_e$ with a Yukawa interaction to a singlet scalar $\phi$ and the left-handed component of a light fermion $\psi$. The Lagrangian of the model is given by
\begin{align}
\mathcal{L}&= -\frac{1}{4}F_{\mu\nu}F^{\mu\nu}+\frac{1}{2}(\partial^\mu \phi)(\partial_\mu \phi)  +\bar \psi\,i \slashed D\, \psi + \bar \Psi(i\slashed{D}-M)\Psi-\left(y\,\bar \psi_L\,\phi\, \Psi_R+\mathrm{h.c.}\right)+\L_\xi,
\end{align}
where $D_\mu\psi=\partial_\mu\psi-ie\, A_\mu\psi$ (similarly for $\Psi$) and $\L_\xi = -(\partial_\mu A^\mu)^2/(2\xi)$ is the gauge-fixing Lagrangian. We illustrate the functional integration of $\Psi$ up to one-loop order and dimension-six operators. The tree-level EFT Lagrangian is easily obtained by substituting the classical value of $\Psi$, defined by its EOM, into the model Lagrangian. The EOM for $\Psi$ reads
\begin{align}\label{eq:PsiEOM}
\Psi&=-\frac{1}{M}\,y^*\,\phi\,\psi_L-\frac{1}{M^2}\,y^*\,i\slashed{D}\,(\phi\,\psi_L)+\mathcal{O}(M^{\eminus 3})\,,
\end{align}
where we ignored terms of $\mathcal{O}(M^{\eminus 3})$, since they do not contribute to the matching of dimension-six operators, neither at tree-level nor at the one-loop order. After substituting~\eqref{eq:PsiEOM} into the model Lagrangian, the tree-level EFT Lagrangian is given by
\begin{align}
\mathcal{L}_\mathrm{\scriptscriptstyle EFT}^\mathrm{\scriptscriptstyle (0)}&=-\frac{1}{4}F_{\mu\nu}F^{\mu\nu}+\frac{1}{2}(\partial^\mu \phi)(\partial_\mu \phi)+\bar \psi\,i \slashed D\, \psi+\L_\xi +\frac{|y|^2}{M^2}\,(\bar\psi_L\,\phi)\,i\slashed{D}\,(\phi\,\psi_L)+\mathcal{O}(M^{\eminus 4})\,.
\end{align}
Let us now proceed to the one-loop matching computation. As discussed in Section~\ref{sec:method}, we fix $\xi=1$ for the quantum fluctuation. Next, we rewrite the fields into multiplets in the form of~\eqref{eq:ComplexForm}:
\begin{align}
\varphi_\phi&=\phi\,,&\varphi_A&=A_\mu\,,&\varphi_\psi&=\begin{pmatrix}\psi\\\psi^c\end{pmatrix}\,,&\varphi_\Psi&=\begin{pmatrix}\Psi\\\Psi^c\end{pmatrix}\,,
\end{align}
with the $c$ superscript denoting charge conjugation. The $X$ terms for this Lagrangian read (cf.~\eqref{eq:XtermExpr})
\begin{align}\label{eq:XVLcase}
X_{\Psi A}^{\scriptscriptstyle [5/2]}&=
\begin{pmatrix}
-e\,\gamma_\mu\,\Psi \\
e\,\gamma_\mu\,\Psi^c
\end{pmatrix}
\,,&
X_{\psi A}^{\scriptscriptstyle [3/2]}&=
\begin{pmatrix}
-e\,\gamma_\mu\,\psi\\
e\,\gamma_\mu\,\psi^c
\end{pmatrix}
\,,&
X_{\psi\Psi}^{\scriptscriptstyle [1]}&=
\begin{pmatrix}
y\, P_R & 0\\
0 & y^*\, P_L
\end{pmatrix}
\phi\,,\nonumber\\
X_{\psi\phi}^{\scriptscriptstyle [7/2]}&=
\begin{pmatrix}
y\, P_R\,\Psi\\
y^*\, P_L\,\Psi^c
\end{pmatrix}
\,,&
X_{\Psi \phi}^{\scriptscriptstyle [3/2]}&=
\begin{pmatrix}
y^*\, P_L\,\psi\\
y\,P_R\,\psi^c
\end{pmatrix}
\,,
\end{align}
while the term with interchanged fields simply correspond to the conjugate of the ones displayed here, e.g. $X_{BA}=\overline{X}_{AB}$. As for the tree-level EFT Lagrangian, the heavy field $\Psi$ needs to be replaced by its classical configuration in~\eqref{eq:PsiEOM}, e.g.
\begin{align}
X_{\psi\phi}^{\scriptscriptstyle [7/2]}&=-\frac{|y|^2}{M^2}\,
\begin{pmatrix}
P_R\,i\slashed{D}\,(\phi\,\psi)\\
P_L\,i\slashed{D}\,(\phi\,\psi^c)
\end{pmatrix}
+\mathcal{O}(M^{-3})\,.
\end{align}
Note that we have added a superindex to the $X$ terms with the total mass dimension of the light fields and covariant derivatives in them. This provides a useful way for counting the mass dimension of a given supertrace. The full set of $X$ terms constitute the main input for \pkg{SuperTracer} to evaluate the one-loop effective Lagrangian.

As described in Section~\ref{sec:method}, the one-loop functional determinant is divided into log-type and power-type contributions, namely $\mathcal{L}_\mathrm{EFT}^{\scriptscriptstyle (1)}=\mathcal{L}_\mathrm{log}^{\scriptscriptstyle (1)}+\mathcal{L}_\mathrm{power}^{\scriptscriptstyle (1)}$. The log-type contribution is obtained from \pkg{SuperTracer} by calling the \rut{LogTerms} routine:
\begin{mmaCell}{Input}
  \mmaDef{LogTerm}[\(\Psi\),6]
\end{mmaCell}
\begin{mmaCell}{Output}
  -\mmaFrac{1}{6} Log\(\Big[\)\mmaFrac{\mmaSup{\(\overline{\mu}\)}{2}}{\mmaSubSup{M}{H}{2}}\(\Big]\)\mmaSup{G}{\(\mu\nu\)}** \mmaSup{G}{\(\mu\nu\)} + \mmaFrac{1}{15}\mmaFrac{1}{\mmaSubSup{M}{H}{2}}\mmaSub{D}{\(\mu\)}\mmaSup{G}{\(\mu\nu\)}** \mmaSub{D}{\(\rho\)}\mmaSup{G}{\(\nu\rho\)} + \mmaFrac{1}{90}i\mmaFrac{1}{\mmaSubSup{M}{H}{2}}\mmaSup{G}{\(\mu\nu\)}** \mmaSup{G}{\(\mu\rho\)}** \mmaSup{G}{\(\nu\rho\)}
\end{mmaCell}
and multiplying the output by $2$ as  $\Psi$ is a Dirac fermion in this example. Since $\Psi$ is charged under an abelian symmetry, we have $ \mathrm{G}_{\mu\nu}=e\,F_{\mu\nu}$, and the resulting Lagrangian reads
\begin{align}
\mathcal{L}_\mathrm{log}^\mathrm{\scriptscriptstyle (1)}&=\frac{e^2}{16\pi^2}\left[-\frac{1}{3}\log\frac{\mu^2}{M_\Psi^2}F_{\mu\nu}F^{\mu\nu}-\frac{2}{15 M_\Psi^2}\,D_\mu F^{\mu\nu} D^\rho F_{\rho\nu}\right]\,.
\end{align}
As described in Section~\ref{subsec:power_traces}, for the power-type contributions, we have to evaluate all possible supertraces constructed out of powers of $\Delta_i\,X_{ij}$ blocks starting with a heavy field propagator, with the sum of $X$ term dimensions not exceeding the desired operator dimension of the EFT Lagrangian. For this example, the power-type Lagrangian up to dimension six is obtained from the following supertraces:
\begin{align}\label{eq:PowerLag}
\int \dd^dx\,\mathcal{L}_\mathrm{power}^{\scriptscriptstyle (1)}&=-\frac{i}{2}\bigg[\mathrm{STr}\left\{\Delta_{\Psi}\,X_{\Psi A}^{\scriptscriptstyle [5/2]}\,\Delta_A\,X_{A\Psi}^{\scriptscriptstyle [5/2]}\right\}+\mathrm{STr}\left\{\Delta_{\Psi}\,X_{\Psi\phi}^{\scriptscriptstyle [3/2]}\,\Delta_\phi\,X_{\phi\Psi}^{\scriptscriptstyle [3/2]}\right\}\nonumber\\
&\quad\left.+\,\mathrm{STr}\left\{\Delta_{\Psi}\,X_{\Psi\psi}^{\scriptscriptstyle [1]}\,\Delta_\psi\,X_{\psi\Psi}^{\scriptscriptstyle [1]}\right\}+\left(\mathrm{STr}\left\{\Delta_{\Psi}\,X_{\Psi A}^{\scriptscriptstyle [5/2]}\,\Delta_A\,X_{A\psi}^{\scriptscriptstyle [3/2]}\,\Delta_{\psi}\,X_{\psi\Psi}^{\scriptscriptstyle [1]}\right\}\right.\right.\nonumber\\
&\quad\left.\left.+\,\mathrm{STr}\left\{\Delta_{\Psi}\,X_{\Psi\psi}^{\scriptscriptstyle [1]}\,\Delta_\psi\,X_{\psi\phi}^{\scriptscriptstyle [7/2]}\,\Delta_{\phi}\,X_{\phi\Psi}^{\scriptscriptstyle [3/2]}\right\}+\mathrm{h.c.}\right)\right.\nonumber\\
&\quad\left.+\,\mathrm{STr}\left\{\Delta_{\Psi}\,X_{\Psi\psi}^{\scriptscriptstyle [1]}\,\Delta_\psi\,X_{\psi A}^{\scriptscriptstyle [3/2]}\,\Delta_{A}\,X_{A\psi}^{\scriptscriptstyle [3/2]}\,\Delta_\psi\,X_{\psi\Psi}^{\scriptscriptstyle [1]}\right\}\right.\nonumber\\
&\quad\left.+\,\mathrm{STr}\left\{\Delta_{\Psi}\,X_{\Psi\psi}^{\scriptscriptstyle [1]}\,\Delta_\psi\,X_{\psi\Psi}^{\scriptscriptstyle [1]}\,\Delta_{\Psi}\,X_{\Psi\phi}^{\scriptscriptstyle [3/2]}\,\Delta_\phi\,X_{\phi\Psi}^{\scriptscriptstyle [3/2]}\right\}\right.\nonumber\\
&\quad\left.+\,\frac{1}{2}\,\mathrm{STr}\left\{\big(\Delta_{\Psi}\,X_{\Psi\phi}^{\scriptscriptstyle [3/2]}\,\Delta_\phi\,X_{\phi\Psi}^{\scriptscriptstyle [3/2]}\big)^2\right\}+\frac{1}{2}\,\mathrm{STr}\left\{\big(\Delta_{\Psi}\,X_{\Psi\psi}^{\scriptscriptstyle [1]}\,\Delta_\psi\,X_{\psi\Psi}^{\scriptscriptstyle [1]}\big)^2\right\}\right.\nonumber\\
&\quad+\,\frac{1}{3}\,\mathrm{STr}\left\{\big(\Delta_{\Psi}\,X_{\Psi\psi}^{\scriptscriptstyle [1]}\,\Delta_\psi\,X_{\psi\Psi}^{\scriptscriptstyle [1]}\big)^3\right\}\bigg]_\mathrm{hard}\,.
\end{align}
The symmetry factors $1/2$ and $1/3$ appearing in front of some of the supertraces count the power of repeated blocks in a given supertrace. The same expression can be readily obtained from \pkg{SuperTracer} with the \rut{PowerTerms} routine:
\begin{mmaCell}{Input}
  Xterms = \{\mmaDef{X}[\{\(\Psi\),\mmaDef{A}\},5/2], \mmaDef{X}[\{\(\psi\),\mmaDef{A}\},3/2], \mmaDef{X}[\{\(\psi\),\(\Psi\)\},1], \mmaDef{X}[\{\(\psi\),\(\phi\)\},7/2],
  \mmaDef{X}[\{\(\Psi\),\(\phi\)\},3/2]\};
  LagPower = \mmaDef{PowerTerms}[Xterms, 6]
\end{mmaCell}
\begin{mmaCell}{Output}
  \mmaDef{STr}[\{\mmaSubSup{X}{\(\Psi\)A}{[5/2]},\mmaSubSup{X}{A\(\Psi\)}{[5/2]}\}] + \mmaDef{STr}[\{\mmaSubSup{X}{\(\Psi\phi\)}{[3/2]},\mmaSubSup{X}{\(\phi\Psi\)}{[3/2]}\}] + \mmaDef{STr}[\{\mmaSubSup{X}{\(\Psi\psi\)}{[1]},\mmaSubSup{X}{\(\psi\Psi\)}{[1]}\}] +
  \mmaDef{STr}[\{\mmaSubSup{X}{\(\Psi\)A}{[5/2]},\mmaSubSup{X}{A\(\psi\)}{[3/2]},\mmaSubSup{X}{\(\psi\Psi\)}{[1]}\}] + \mmaDef{STr}[\{\mmaSubSup{X}{\(\Psi\phi\)}{[3/2]},\mmaSubSup{X}{\(\phi\psi\)}{[7/2]},\mmaSubSup{X}{\(\psi\Psi\)}{[1]}\}] +
  \mmaDef{STr}[\{\mmaSubSup{X}{\(\Psi\psi\)}{[1]},\mmaSubSup{X}{\(\psi\)A}{[3/2]},\mmaSubSup{X}{A\(\Psi\)}{[5/2]}\}] +  \mmaDef{STr}[\{\mmaSubSup{X}{\(\Psi\psi\)}{[1]},\mmaSubSup{X}{\(\psi\phi\)}{[7/2]},\mmaSubSup{X}{\(\phi\Psi\)}{[3/2]}\}] +
  \mmaDef{STr}[\{\mmaSubSup{X}{\(\Psi\phi\)}{[3/2]},\mmaSubSup{X}{\(\phi\Psi\)}{[3/2]},\mmaSubSup{X}{\(\Psi\phi\)}{[3/2]},\mmaSubSup{X}{\(\phi\Psi\)}{[3/2]}\}] + \mmaDef{STr}[\{\mmaSubSup{X}{\(\Psi\psi\)}{[1]},\mmaSubSup{X}{\(\psi\)A}{[3/2]},\mmaSubSup{X}{A\(\psi\)}{[3/2]},\mmaSubSup{X}{\(\psi\Psi\)}{[1]}\}] + 
  \mmaDef{STr}[\{\mmaSubSup{X}{\(\Psi\psi\)}{[1]},\mmaSubSup{X}{\(\psi\Psi\)}{[1]},\mmaSubSup{X}{\(\Psi\phi\)}{[3/2]},\mmaSubSup{X}{\(\phi\Psi\)}{[3/2]}\}] + \mmaDef{STr}[\{\mmaSubSup{X}{\(\Psi\psi\)}{[1]},\mmaSubSup{X}{\(\psi\Psi\)}{[1]},\mmaSubSup{X}{\(\Psi\psi\)}{[1]},\mmaSubSup{X}{\(\psi\Psi\)}{[1]}\}] + 
  \mmaDef{STr}[\{\mmaSubSup{X}{\(\Psi\psi\)}{[1]},\mmaSubSup{X}{\(\psi\Psi\)}{[1]},\mmaSubSup{X}{\(\Psi\psi\)}{[1]},\mmaSubSup{X}{\(\psi\Psi\)}{[1]},\mmaSubSup{X}{\(\Psi\psi\)}{[1]},\mmaSubSup{X}{\(\psi\Psi\)}{[1]}\}]
\end{mmaCell}
This routine takes as input all $X$ terms present in a given model, defined in \mmaInlineCell[]{Code}{Xterms} and the maximal mass dimension of the supertraces, which is \mmaInlineCell[]{Code}{6} in our example. The routine \rut{PowerTerms} automatically completes the list of $X$ interactions in \mmaInlineCell[]{Code}{Xterms} with the corresponding conjugate interactions, namely $X_{A\psi}^{\scriptscriptstyle [5/2]}$, $X_{A\psi}^{\scriptscriptstyle [3/2]}$, etc. are automatically included, so their input is optional. Further note that the $-\frac{i}{2}$ and symmetry factors are absorbed in the definition of \rut{STr} in \pkg{SuperTracer}. Moreover, the field propagators in between $X$ terms are implicitly understood. These supertraces can be evaluated by replacing \rut{STr} by \rut{STrTerms}, while the loop functions are evaluated using the \rut{EvaluateLoopFunctions} routine. For concreteness, let us focus in the first term of this expression:
\begin{mmaCell}{Input}
  \mmaDef{LagPower}[[1]]
  \% /. \mmaDef{STr} -> \mmaDef{STrTerm}
  \%//\mmaDef{EvaluateLoopFunctions}//\mmaDef{SuperSimplify}
\end{mmaCell}
\begin{mmaCell}{Output}
  \mmaDef{STr}[\{\mmaSubSup{X}{\(\Psi\)A}{[5/2]},\mmaSubSup{X}{A\(\Psi\)}{[5/2]}\}]
\end{mmaCell}
\begin{mmaCell}{Output}
  \mmaFrac{1}{8}i\(\big(\)1 + 2 \mmaSub{\mmaDef{LF}}{1,1}[\mmaDef{\mmaSub{M}{H}}]\(\big)\)\mmaSub{\(\gamma\)}{\(\mu\)}**\mmaSub{D}{\(\mu\)}\mmaSubSup{X}{\mmaSub{\(\Psi\)}{i}\mmaSub{A}{j}} **\mmaSubSup{X}{\mmaSub{\mmaSub{A}{j}\(\Psi\)}{i}} +\mmaFrac{1}{2} \mmaSub{\mmaDef{LF}}{1,1}[\mmaDef{\mmaSub{M}{H}}] \mmaSub{M}{H}\mmaSubSup{X}{\mmaSub{\(\Psi\)}{i}\mmaSub{A}{j}} **\mmaSubSup{X}{\mmaSub{\mmaSub{A}{j}\(\Psi\)}{i}}
\end{mmaCell}
\begin{mmaCell}{Output}
  \mmaFrac{1}{8}i\(\bigg(\)3 + 2 Log\(\Big[\)\mmaFrac{\mmaSup{\(\overline{\mu}\)}{2}}{\mmaSubSup{M}{H}{2}}\(\Big]\)\(\bigg)\)\mmaSub{\(\gamma\)}{\(\mu\)}**\mmaSub{D}{\(\mu\)}\mmaSubSup{X}{\mmaSub{\(\Psi\)}{i}\mmaSub{A}{j}} **\mmaSubSup{X}{\mmaSub{\mmaSub{A}{j}\(\Psi\)}{i}} +\mmaFrac{1}{2}\(\bigg(\)1 + Log\(\Big[\)\mmaFrac{\mmaSup{\(\overline{\mu}\)}{2}}{\mmaSubSup{M}{H}{2}}\(\Big]\)\(\bigg)\)\mmaSub{M}{H}\mmaSubSup{X}{\mmaSub{\(\Psi\)}{i}\mmaSub{A}{j}} **\mmaSubSup{X}{\mmaSub{\mmaSub{A}{j}\(\Psi\)}{i}}
\end{mmaCell}
Note that, for notational simplicity, $\int\, \dd^d x\,\frac{1}{16\pi^2}$ is omitted in the output of \rut{STrTerms}. 

The \pkg{SuperTracer} package also allows for the substitution of the $X$ interactions. Let us continue to use the above term as an example. First, we need to define the fields appearing in a given $X$ term. Since in this case we want to replace $X_{\Psi A}$ in~\eqref{eq:XVLcase}, we only need to define the heavy fermion field $\Psi$. This is done by calling the routine \rut{AddField}:
\begin{mmaCell}{Input}
  \mmaDef{AddField}[\mmaUnd{\(\psi\)}h, \(\Psi\), e[1]]
\end{mmaCell}
where the first argument is the label we are going to use for the field, the second argument is the type of field, and the third argument the $\U(1)_e$ charge. In this example, we are denoting the $\Psi$ field with the label `$\psi_h$,' and we are defining it as a heavy fermion field, $\Psi$ in \pkg{SuperTracer} notation. Note that we have avoided using the label `$\Psi$', since this variable is already predefined in \pkg{SuperTracer}. Once the field has been defined, we can introduce the value of $X_{\Psi A}$ in~\eqref{eq:XVLcase} (and its conjugate) into the \rut{STrTerm} routine:
\begin{mmaCell}{Input}
  \mmaDef{STrTerm}[\{\mmaDef{X}[\{\(\Psi\),\mmaDef{A}\},5/2],\mmaDef{X}[\{\mmaDef{A},\(\Psi\)\},5/2]\},6,
     \{
      \{\(\Psi\),\mmaDef{A}\}->\{\{-e \(\gamma\)[\mmaUnd{\(\alpha\)}[j]]**\mmaDef{\(\psi\)h[]}\}, \{e \(\gamma\)[\mmaUnd{\(\alpha\)}[j]]**\mmaDef{CConj}[\mmaDef{\(\psi\)h[]]}\}\},
      \{\mmaDef{A},\(\Psi\)\}->\{\{-e \mmaDef{Bar}[\mmaDef{\(\psi\)h}[]]**\(\gamma\)[\mmaUnd{\(\alpha\)}[i]], e \mmaDef{Bar}[\mmaDef{CConj}[\mmaDef{\(\psi\)h[]]}]**\(\gamma\)[\mmaUnd{\(\alpha\)}[i]]\}\},
      \mmaDef{M}[\(\Psi\)]->\{\mmaUnd{Mh}, \mmaUnd{Mh}\},
      \mmaDef{G}[\(\Psi\)]->\{\{e[1]\},\{e[-1]\}\},
      \mmaDef{G}[\mmaDef{A}]->\{\{\}\}
     \}
   ]//\mmaDef{SuperSimplify}//\mmaDef{EvaluateLoopFunctions}
\end{mmaCell}
\begin{mmaCell}{Output}
  \(\,\)\mmaFrac{1}{2}\textbf{}i\(\,\)\mmaSup{e}{2}\(\bigg(\)1 + 2 Log\(\Big[\)\mmaFrac{\mmaSup{\(\overline{\mu}\)}{2}}{\mmaSup{Mh}{2}}\(\Big]\)\(\bigg)\)\mmaDef{\(\overline{\psi\mathrm{h}}\)}**\mmaSub{\(\gamma\)}{\(\mu\)}**\mmaSub{D}{\(\mu\)}\(\psi\mathrm{h}\) - 2\(\,\)\mmaSup{e}{2}\(\,\)Mh\(\bigg(\)1 + 2 Log\(\Big[\)\mmaFrac{\mmaSup{\(\overline{\mu}\)}{2}}{\mmaSup{Mh}{2}}\(\Big]\)\(\bigg)\)\mmaDef{\(\overline{\psi\mathrm{h}}\)}**\mmaDef{\(\psi\)h}
\end{mmaCell}
A few comments on the notation are in order:
\begin{enumerate}[i)]
    \item Substitution rules that are not scalar have to be introduced in matrix form. The substitution for $\{\Psi,A\}$ is a column vector (\{\{a\}\},\{b\}\}) and for $\{A,\Psi\}$ a row vector (\{\{a,b\}\}), corresponding to $ \varphi_A $ being a scalar and $ \varphi_\Psi $ being a doublet. 
    
    \item The use of \mmaInlineCell[]{Input}{NonCommutativeMultiply} (denoted by \mmaInlineCell[]{Input}{**})  when multiplying fields is mandatory, since these need to be treated as non-commuting objects during \pkg{SuperTracer} evaluation.
    
    \item Whenever there are vector fields in the substitution rules, there needs to be an open Lorentz index matching that of the vector field. This Lorentz index has to be \mmaInlineCell[]{Input}{\mmaUnd{\(\alpha\)}[i]} when $A$ is the first element, e.g $\{A,\Psi\}$, and \mmaInlineCell[]{Input}{\mmaUnd{\(\alpha\)}[j]} when $A$ is the second element, as in $\{\Psi,A\}$. 
    
    \item We have also defined the heavy fermion masses with the third substitution rule. We have avoided using \mmaInlineCell[]{Input}{\mmaDef{M}} for the mass, since this variable is already predefined in \pkg{SuperTracer}. Also, since the heavy fermion field is encoded in the $ \varphi_\Psi = (\Psi\;\Psi^c)^\intercal$ doublet, a list with two elements is needed.
    
    \item Finally, we have defined the action of the field-strength tensors on the fields. In this case, there is a single gauge group, the $\U(1)_e$, which we labeled with \mmaInlineCell[]{Input}{e}, so we only need to specify the electric charges in the format \mmaInlineCell[]{Input}{\{e[charge]\}} for each of the fields. By default, the charges are assumed to be zero, which is why we input an empty list for $A$. A more complicated example with multiple gauge groups is given in the next section.
\end{enumerate}
After substituting the EOM into the output  for $\Psi$ (see~\eqref{eq:PsiEOM}), one readily obtains\footnote{As we show in the ancillary \pkg{Mathematica} notebook \textit{VLfermExample.nb}, it is also possible to include the EOM for $\Psi$ in the input of \rut{STrTerm}, yielding the same result.}
\begin{align}
{\color{blue}-\frac{i}{2}}\,\mathrm{STr}\left\{{\color{blue}\Delta_{\Psi}}\,X_{\Psi A}^{\scriptscriptstyle [5/2]}\,{\color{blue}\Delta_A}\,X_{A\Psi}^{\scriptscriptstyle [5/2]}\right\}_{\color{blue}\mathrm{hard}}=-{\color{blue}\int \dd^dx\,\frac{1}{16\pi^2}}\,7e^2\left(\frac{1}{2}+\log\frac{\bar\mu^2}{M_\Psi^2}\right)\!\frac{|y|^2}{M_\Psi^2}\,(\bar\psi_L\,\phi)\,i\slashed{D}\,(\phi\,\psi_L)\,,
\end{align}
where the parts highlighted in {\color{blue}blue} are kept implicit in \pkg{SuperTracer} for notational simplicity. The complete computation of the power-type Lagrangian is provided in the ancillary \pkg{Mathematica} notebook \textit{VLfermExample.nb}. We have compared this result against an explicit computation done by diagrammatic matching, finding full agreement between the two. More details on this comparison are provided in Appendix~\ref{app:VLexample}.

We wish to close this section with a consequence of the $\gamma_5$ prescription employed in our approach. The supertrace
\begin{align}
 \mathrm{STr}\left\{\Delta_{\Psi}\,X_{\Psi\psi}^{\scriptscriptstyle [1]}\,\Delta_\psi\,X_{\psi\Psi}^{\scriptscriptstyle [1]}\right\}\,,
\end{align}
contains terms with divergent loop integrals and odd numbers of $\gamma_5$. Due to the lack of a CP-violating interactions in the model, these terms cannot give rise to a contribution to the effective action of the form $F_{\mu\nu}\tilde F^{\mu\nu}\phi^2$. Indeed, in the \pkg{SuperTracer} calculation, a cancellation between the contributions from $\psi,\Psi$ and the ones from $\psi^c,\Psi^c$ takes place. This result is found in our prescription only if the traces are read from the correct starting point, which is guaranteed by construction in our formalism. At the diagrammatic level, a reading point ambiguity persists unless the diagrams are read in a consistent way. This means that some Dirac traces would have to be read against the conventional direction and interpreted as loops of charge-conjugated fermions instead of the usual way. We stress again that in our approach, the traces are automatically arranged in a way that fixed this issue. We have also checked against the diagrammatic computation that the WC of an operator of the form $F_{\mu\nu}\tilde F^{\mu\nu}\phi_1\,\phi_2$, in a theory with two scalar fields instead of one, is correctly reproduced.

\subsection{\texorpdfstring{$S_1$}{S1} scalar leptoquark}
\label{subsec:S1example}

As our second example, we consider an $S_1\sim(\mathbf{\bar 3},\mathbf{1},1/3)$ scalar leptoquark extension of the SM, with the parenthesis indicating the $S_1$ representation under the SM gauge group $\SU(3)_c\times \SU(2)_L\times \U(1)_Y$. The Lagrangian of the model reads
\begin{align}
\mathcal{L}&=\mathcal{L}_\mathrm{SM}\,+|D_\mu S_1|^2-M^2\,|S_1|^2-\big(\lambda_{1L}^{i\alpha}\, \bar q^c_i\epsilon\,\ell_\alpha\,S_1+\lambda_{1R}^{i\alpha}\,\bar u^c_i\,e_\alpha\,S_1+\mathrm{h.c.}\big) \nonumber\\
&\quad-\frac{\lambda_S}{2}\,|S_1|^4-\lambda_{H S}\,|H|^2\,|S_1|^2\,,
\end{align}
where $\mathcal{L}_\mathrm{SM}$ is the SM Lagrangian, $\epsilon=i\,\sigma_2$ is the $\SU(2)_L$ anti-symmetric tensor, and $i$ and $\alpha$ are quark and lepton flavor indices, respectively. The covariant derivative acting on $S_1$ is given by 
\begin{align}
D_\mu S_1=\left(\partial_\mu+ig_c\, (T^a)^* G_\mu^{a}-\frac{1}{3}\,ig_Y B_\mu\right)\,S_1\,,
\end{align}
with $T^a$ being the fundamental $\SU(3)$ generators, and $g_c$ and $g_Y$ the QCD and hypercharge gauge couplings, respectively. The complete one-loop matching conditions of this model to the SMEFT up to dimension-six operators can be found in~\cite{Gherardi:2020det}. Here, we do not intend to fully reproduce this result but rather to illustrate the one-loop matching procedure using the functional method described in Section~\ref{sec:method} and the \pkg{SuperTracer} package. First, we obtain the tree-level effective Lagrangian by substituting the EOM of $S_1$,
\begin{align}
S_1=\frac{1}{M^2}\,\left[(\lambda_{1L}^{i\alpha})^*\, \bar\ell_\alpha\, \epsilon\,q_i^c\,-(\lambda_{1R}^{i\alpha})^*\,\bar e_\alpha u_i^c\right]+\mathcal{O}(M^{\eminus 2})\,,
\end{align}
into the Lagrangian, yielding
\begin{align}
\mathcal{L}=&\,\mathcal{L}_\mathrm{SM}\,
-\frac{1}{M^2}\,\lambda_{1L}^{i\alpha}(\lambda_{1L}^{j\beta})^*\, (\bar q^c_i\,\epsilon\,\ell_\alpha)\,(\bar\ell_\beta\,\epsilon\, q_j^c)
+\frac{1}{M^2}\,\lambda_{1R}^{i\alpha}(\lambda_{1R}^{j\beta})^*\, (\bar u^c_i\,e_\alpha)\,(\bar e_\beta\, u_j^c)\nonumber \\
&+\frac{1}{M^2}\big[\lambda_{1L}^{i\alpha}(\lambda_{1R}^{j\beta})^*\, (\bar q^c_i\,\epsilon\,\ell_\alpha)\,(\bar e_\beta \,u_j^c)+\mathrm{h.c.}\big)+\mathcal{O}(M^{\eminus 4})\,,
\end{align}
which after applying Fierz transformations coincides with the tree-level Lagrangian in~\cite{Gherardi:2020det}. To perform the one-loop integration, we collect the fields into multiplets in the form of~\eqref{eq:ComplexForm}:
\begin{align}
\varphi_S&=\begin{pmatrix}S_1\\ S_1^*\end{pmatrix}\,,&
\varphi_H&=\begin{pmatrix}H\\ H^*\end{pmatrix}\,,&
\varphi_f&=\begin{pmatrix}f\\ f^c\end{pmatrix}\,,&
\varphi_A&=A\,,&
\end{align}
with $A=B,W,G$ and $f=q,u,d,\ell,e$. As in the previous example, the $X$ terms for the $S_1$ part of the Lagrangian can be readily obtained from~\eqref{eq:XtermExpr}:
\begin{align}\label{eq:S1Xterms}
X_{SA}^{\scriptscriptstyle[4,3]}&=
-\begin{pmatrix}
2i\,Q_{SA}\,(D_\mu\,S_1)\\
-2i\,Q_{SA}^*\,(D_\mu\,S_1)^*
\end{pmatrix}
-
\begin{pmatrix}
Q_{SA}\,S_1\\
-Q_{SA}^*\,S_1^*
\end{pmatrix}
iD_\mu 
\,, \span
\span \quad 
X_{AA'}^{\scriptscriptstyle[6]}= - g_{\mu\nu} S_1^\dagger \big\{Q_{SA}, Q_{SA'} \big\} S_1\,,
\nonumber\\
X_{qS}^{\scriptscriptstyle[3/2]}&=
\begin{pmatrix}
0 & \lambda_{1L}^*\,\epsilon\, P_R \ell^c\\
\lambda_{1L}\,\epsilon\,P_L \ell &  0
\end{pmatrix}
\,, & \qquad
X_{\ell S}^{\scriptscriptstyle[3/2]} &=
\begin{pmatrix}
0 & -\lambda_{1L}^*\,\epsilon\,P_R\,q^c\\
-\lambda_{1L}\,\epsilon\,P_L\,q & 0
\end{pmatrix}
\,,\nonumber\\
X_{uS}^{\scriptscriptstyle[3/2]}&=
\begin{pmatrix}
0 & \lambda_{1R}^* \, P_L\,e^c \\
\lambda_{1R} \,P_R\,e & 0
\end{pmatrix}
\,, & \qquad
X_{e S}^{\scriptscriptstyle[3/2]} &=
\begin{pmatrix}
0 & \lambda_{1R}^*\,P_L\, u^c\\
\lambda_{1R}\,P_R\, u & 0
\end{pmatrix}
\,,\nonumber\\
X_{ql}^{\scriptscriptstyle[3]}&=
\begin{pmatrix}
0 & \lambda_{1L}^* \epsilon\,P_R S_1^*\\
 \lambda_{1L} \epsilon\, P_L S_1 & 0
\end{pmatrix} 
\,, & \qquad
X_{ue}^{\scriptscriptstyle[3]} &=
\begin{pmatrix}
0 & \lambda_{1R}^* P_L S_1^*\\
\lambda_{1R} P_R S_1 & 0
\end{pmatrix}
\,,\nonumber\\
X_{SH}^{\scriptscriptstyle[4]}&=\lambda_{HS}\begin{pmatrix}
S_1 H^\dagger & S_1 H^\intercal\\
S_1^* H^\dagger & S_1^* H^\intercal
\end{pmatrix}
\,, & \qquad
X_{HH}^{\scriptscriptstyle[6]} &= \lambda_{HS}
\begin{pmatrix}
(S_1^\dagger S_1)\,\mathbb{1} & 0\\
0 & (S_1^\dagger S_1)\,\mathbb{1}
\end{pmatrix},\nonumber\\
X_{SS}^{\scriptscriptstyle[2]}&=
\lambda_{HS}
\begin{pmatrix}
(H^\dagger H)\,\mathbb{1} &0\\
0 &  (H^\dagger H)\,\mathbb{1}
\end{pmatrix}+\lambda_{S}
\begin{pmatrix}
  (S_1^\dagger S_1)\,\mathbb{1} + S_1 S_1^\dagger &  S_1S_1^\intercal\\
 S_1^*S_1^\dagger & (S_1^\dagger S_1)\,\mathbb{1} + S_1 S_1^\dagger 
\end{pmatrix},\span\span 
\end{align}
with $Q_{SB}=g^\prime/3$, $Q_{SW}=0$ and $Q_{SG}= -g_s\, (T^a)^*$. The corresponding $X$ terms for the SM interactions can be found e.g. in Appendix B of~\cite{Cohen:2020fcu}. As in the previous example, all the $X$ terms with permutated fields can be obtained by Hermitian conjugation of the ones above, that is $X_{BA}=\overline{X}_{AB}$. 
However, in contrast with the previous example, we now have an ``open covariant derivative'', i.e. a covariant derivative that does not act inside a commutator, in the $X_{SA}$ interaction. Following the prescription in~\eqref{eq:Xexpansion}, this means that $X_{AS}$ should be put in canonical form by making the derivative act from the rightmost, e.g.
\begin{align}\label{eq:XAS}
X_{AS}^{\scriptscriptstyle[4,3]}&=
\begin{pmatrix}
i(D_\mu\,S_1)^\dagger\, Q_{SA} &
-i(D_\mu\,S_1)^\intercal\,Q_{SA}^*
\end{pmatrix}
+
\begin{pmatrix}
-S_1^\dagger\, Q_{SA} &
S_1^\intercal\,Q_{SA}^*
\end{pmatrix}
iD_\mu
\,.
\end{align}
We have once again included the mass dimension of the $X$ terms as a superscript. For the $X_{SA}$ and $X_{AS}$, which contain $X_0$ and $X_1^\mu$ terms in the expansion in~\eqref{eq:Xexpansion}, we have added two counting parameters instead of one, corresponding respectively to the term without open derivatives, $X_0$, and the term with one open derivative, $X_1^\mu$. The open derivative is not included in the counting of the $X$ mass dimensions. We emphasize that specifying $X$ mass dimensions in this way is useful to keep track of the EFT power counting. 

Once the $X$ terms have been determined, we can proceed to the identification and evaluation of the relevant log-type and power-type supertraces yielding $\mathcal{L}_\mathrm{EFT}^{\scriptscriptstyle (1)}=\mathcal{L}_\mathrm{log}^{\scriptscriptstyle (1)}+\mathcal{L}_\mathrm{power}^{\scriptscriptstyle (1)}$. Once more, the log-type contribution can be readily obtained:
\begin{align}\label{eq:S1LogTerm}
\mathcal{L}_\mathrm{log}^{\scriptscriptstyle(1)}&=\frac{1}{16\pi^2}\left[-\frac{1}{12}\log\frac{\mu^2}{M^2}\,\mathrm{tr}_G\left\{F_{\mu\nu}F^{\mu\nu}\right\}-\frac{1}{60 M^2}\,\mathrm{tr}_G\left\{D_\mu F^{\mu\nu} D^\rho F_{\rho\nu}\right\}\right.\nonumber\\
&\quad\left.-\frac{1}{90 M^2}\,\mathrm{tr}_G\left\{i\,F_\mu^{\;\;\nu}F_\nu^{\;\;\rho}F_\rho^{\;\;\mu}\right\}\right]\nonumber\\
&=\frac{1}{16\pi^2}\left[-\frac{g_c^2}{24}\log\frac{\mu^2}{M^2}\,(G_{\mu\nu})^a(G^{\mu\nu})^a-\frac{g_Y^2}{36}\log\frac{\mu^2}{M^2}\,B_{\mu\nu}B^{\mu\nu}-\frac{g_c^2}{120 M^2}\,(D_\mu G^{\mu\nu})^a(D^\rho G_{\rho\nu})^a\right.\nonumber\\
&\quad\left.\!-\frac{g_Y^2}{180 M^2}\,\partial_\mu B^{\mu\nu} \partial^\rho B_{\rho\nu}+\frac{g_c^3}{360 M^2}\,f_{abc}\,(G_\mu^{\;\;\nu})^a(G_\nu^{\;\;\rho})^b(G_\rho^{\;\;\mu})^c\right],
\end{align}
where we took $F_{\mu\nu}=g_c\,T^a\,(G_{\mu\nu})^a+g_Y\,Y_{S_1}\,B_{\mu\nu}$ in the second equality. Note the implicit color factor arising from the gauge trace in the terms with $B_{\mu\nu}$. This result coincides with the one in~\cite{Gherardi:2020det}. The same expression for the first equality in~\eqref{eq:S1LogTerm} is obtained by \pkg{SuperTracer} by running 
\begin{mmaCell}{Input}
  \mmaDef{LogTerm}[\(\Phi\),6]
\end{mmaCell}
\begin{mmaCell}{Output}
  -\mmaFrac{1}{24} Log\(\Big[\)\mmaFrac{\mmaSup{\(\overline{\mu}\)}{2}}{\mmaSubSup{M}{H}{2}}\(\Big]\)\mmaSup{G}{\(\mu\nu\)}** \mmaSup{G}{\(\mu\nu\)} + \mmaFrac{1}{120}\mmaFrac{1}{\mmaSubSup{M}{H}{2}}\mmaSub{D}{\(\mu\)}\mmaSup{G}{\(\mu\nu\)}** \mmaSub{D}{\(\rho\)}\mmaSup{G}{\(\nu\rho\)} - \mmaFrac{1}{180}i\mmaFrac{1}{\mmaSubSup{M}{H}{2}}\mmaSup{G}{\(\mu\nu\)}** \mmaSup{G}{\(\mu\rho\)}** \mmaSup{G}{\(\nu\rho\)}
\end{mmaCell}
and accounting for the doubling of contributions since $S_1$ is a complex scalar field, and the contributions from both $S_1$ and $S_1^*$ should be included. As for the power terms, the first thing to note is that $X_{AA}$, $X_{SH}$ and $ X_{HH}$ do not contribute at mass dimension six due to their high mass dimension. Since we do not intend to perform the full matching procedure, but just to illustrate the method in a more realistic example, we set $\lambda_{1R}$ to zero and neglect the SM Yukawa couplings. In this case, the only relevant SM $X$ terms are $X_{\psi A}^{\scriptscriptstyle[3/2]}$ and their conjugates. Following the prescription in Section~\ref{subsec:summaryPI}, we collect all fields of the same type into multiplets, such that e.g. $\varphi_A=(G \; W \; B)^\intercal$ and $X_{\Phi A}=(X_{SG} \; 0 \; S_{SB})$. The Lagrangian for the power terms then reads
\begin{align}
\mathcal{L}_\mathrm{power}^{\scriptscriptstyle (1)}=-&\frac{i}{2}\bigg[
\mathrm{STr}\left\{\Delta_\Phi\,X_{\Phi\Phi}^{\scriptscriptstyle[2]}\right\}
+\mathrm{STr}\left\{\Delta_\Phi\,X_{\Phi A}^{\scriptscriptstyle[4,3]}\,\Delta_A\,X_{A\Phi}^{\scriptscriptstyle [4,3]}\right\}
+\frac{1}{2}\,\mathrm{STr}\left\{\left(\Delta_\Phi\,X_{\Phi\Phi}^{\scriptscriptstyle[2]}\right)^2\right\}
 \nonumber \\
&\left. +\mathrm{STr} \left\{\Delta_\Phi X_{\Phi\psi}^{\scriptscriptstyle[3/2]} \Delta_\psi X_{\psi \Phi}^{\scriptscriptstyle[3/2]}\right\}
+\frac{1}{3}\,\mathrm{STr}\left\{\left(\Delta_\Phi\,X_{\Phi\Phi}^{\scriptscriptstyle[2]}\right)^3\right\} 
\right. \nonumber \\
& \left.
+ \mathrm{STr}\left\{ \Delta_\Phi X_{\Phi\Phi}^{\scriptscriptstyle[2]} \Delta_\Phi X_{\Phi\psi}^{\scriptscriptstyle[3/2]}\Delta_\psi X_{\psi \Phi}^{\scriptscriptstyle[3/2]}\right\}
+\left(\mathrm{STr}\left\{ \Delta_\Phi X_{\Phi A}^{\scriptscriptstyle[4,3]}\Delta_A X_{A\psi}^{\scriptscriptstyle[3/2]}\Delta_\psi X_{\psi \Phi}^{\scriptscriptstyle[3/2]} \right\}
+\mathrm{h.c.}\right)
 \right.\nonumber \\
&\left.
+\mathrm{STr} \left\{ \Delta_\Phi X_{\Phi\psi}^{\scriptscriptstyle[3/2]}\Delta_\psi X_{\psi\psi}^{\scriptscriptstyle[3]} \Delta_\psi X_{\psi \Phi}^{\scriptscriptstyle[3/2]} \right\}
+\mathrm{STr}\left\{ \Delta_\Phi X_{\Phi\psi}^{\scriptscriptstyle[3/2]} \Delta_\psi X_{\psi A}^{\scriptscriptstyle[3/2]}\Delta_A X_{A\psi}^{\scriptscriptstyle[3/2]} \Delta_\psi X_{\psi \Phi}^{\scriptscriptstyle[3/2]} \right\}
 \right.\nonumber \\ 
&+ \frac{1}{2}\,\mathrm{STr} \left\{ \left( \Delta_\Phi X_{\Phi\psi}^{\scriptscriptstyle[3/2]} \Delta_\psi X_{\psi \Phi}^{\scriptscriptstyle[3/2]} \right)^2 \right\}
\bigg]_\mathrm{hard}\,.
\end{align}
Once more, note the symmetry factors $1/2$ and $1/3$ in some of the traces. This result is reproduced by the program from the input
\begin{mmaCell}{Input}
  Xterms = \{\mmaDef{X}[\{\(\Phi\),\mmaDef{A}\},\{4,3\}],\mmaDef{X}[\{\(\psi\),\(\Phi\)\},3/2],\mmaDef{X}[\{\(\psi\),\(\psi\)\},3],
            \mmaDef{X}[\{\(\Phi\),\(\Phi\)\},2],\mmaDef{X}[\{\(\psi\),\mmaDef{A}\},3/2]\};
  LagPower = \mmaDef{PowerTerms}[Xterms]
\end{mmaCell}
\begin{mmaCell}{Output}
 \mmaDef{STr}[\{\mmaSubSup{X}{\(\Phi\)\(\Phi\)}{[2]}\}] + \mmaDef{STr}[\{\mmaSubSup{X}{\(\Phi\)\mmaDef{A}}{[\{4,3\}]},\mmaSubSup{X}{\mmaDef{A}\(\Phi\)}{[\{4,3\}]}\}] + \mmaDef{STr}[\{\mmaSubSup{X}{\(\Phi\)\(\Phi\)}{[2]},\mmaSubSup{X}{\(\Phi\)\(\Phi\)}{[2]}\}]
  + \mmaDef{STr}[\{\mmaSubSup{X}{\(\Phi\)\(\psi\)}{[3/2]},\mmaSubSup{X}{\(\psi\)\(\Phi\)}{[3/2]}\}] + \mmaDef{STr}[\{\mmaSubSup{X}{\(\Phi\)\mmaDef{A}}{[\{4,3\}]},\mmaSubSup{X}{\mmaDef{A}\(\psi\)}{[3/2]},\mmaSubSup{X}{\(\psi\)\(\Phi\)}{[3/2]}\}] 
  + \mmaDef{STr}[\{\mmaSubSup{X}{\(\Phi\)\(\Phi\)}{[2]},\mmaSubSup{X}{\(\Phi\)\(\Phi\)}{[2]},\mmaSubSup{X}{\(\Phi\)\(\Phi\)}{[2]}\}] + \mmaDef{STr}[\{\mmaSubSup{X}{\(\Phi\)\(\Phi\)}{[2]},\mmaSubSup{X}{\(\Phi\)\(\psi\)}{[3/2]},\mmaSubSup{X}{\(\psi\)\(\Phi\)}{[3/2]}\}] 
  + \mmaDef{STr}[\{\mmaSubSup{X}{\(\Phi\)\(\psi\)}{[3/2]},\mmaSubSup{X}{\(\psi\)\mmaDef{A}}{[3/2]},\mmaSubSup{X}{\mmaDef{A}\(\Phi\)}{[\{4,3\}]}\}] + \mmaDef{STr}[\{\mmaSubSup{X}{\(\Phi\)\(\psi\)}{[3/2]},\mmaSubSup{X}{\(\psi\)\(\psi\)}{[3]},\mmaSubSup{X}{\(\psi\)\(\Phi\)}{[3/2]}\}] 
  + \mmaDef{STr}[\{\mmaSubSup{X}{\(\Phi\)\(\psi\)}{[3/2]},\mmaSubSup{X}{\(\psi\)\mmaDef{A}}{[3/2]},\mmaSubSup{X}{\mmaDef{A}\(\psi\)}{[3/2]},\mmaSubSup{X}{\(\psi\)\(\Phi\)}{[3/2]}\}] + \mmaDef{STr}[\{\mmaSubSup{X}{\(\Phi\)\(\psi\)}{[3/2]},\mmaSubSup{X}{\(\psi\)\(\Phi\)}{[3/2]},\mmaSubSup{X}{\(\Phi\)\(\psi\)}{[3/2]},\mmaSubSup{X}{\(\psi\)\(\Phi\)}{[3/2]}\}]
\end{mmaCell}
where, we remind the reader, the symmetry and the $-i/2$ factors are taken as part of the definition of \rut{STr} in \pkg{SuperTracer}. As an example, we show the evaluation of the second term, corresponding in the diagrammatic language to a one-loop gauge correction to the propagator of the $S_1$ leptoquark. As usual, first we have to define the fields entering in the $X$ substitutions using \rut{AddField}, which in this case is just the  $S_1$ leptoquark:
\begin{mmaCell}{Input}
  \mmaDef{AddField}[S1,\(\Phi\),Y[1/3]]
\end{mmaCell}
where we labeled hypercharge by \mmaInlineCell[]{Input}{Y}. Once this has been done, we can input our $X$ substitution in~\eqref{eq:S1Xterms} into \rut{STrTerm} to obtain desired result:
\begin{mmaCell}{Input}
  \mmaDef{STrTerm}[\{\mmaDef{X}[\{\(\Phi\),\mmaDef{A}\}, \{4,3\}], \mmaDef{X}[\{\mmaDef{A},\(\Phi\)\}, \{4,3\}]\}, 6, 
   \{ 
    \{\(\Phi\),\mmaDef{A}\} -> - \mmaDef{g}[\mmaUnd{\(\alpha\)}[j], \mmaUnd{\(\mu\)}] 
     \{
      \{-gc \mmaDef{Bar}[\mmaDef{T}[\{SU3A[j], SU3[i], SU3[a]\}]] \mmaDef{S1}[\{SU3[a]\}],0,
        gp/3 \mmaDef{S1}[\{SU3[i]\}]\},
      \{gc \mmaDef{T}[\{SU3A[j], SU3[i], SU3[a]\}] \mmaDef{Bar}[\mmaDef{S1}[\{SU3[a]\}]],0,
       -gp/3 \mmaDef{Bar}[\mmaDef{S1}[\{SU3[i]\}]]\}
     \},
    \{\mmaDef{A},\(\Phi\)\} -> - \mmaDef{g}[\mmaUnd{\(\alpha\)}[i], \mmaUnd{\(\mu\)}] 
     \{
      \{-gc \mmaDef{T}[\{SU3A[i], SU3[j], SU3[a]\}] \mmaDef{Bar}[\mmaDef{S1}[\{SU3[a]\}]],
         gc \mmaDef{Bar}[\mmaDef{T}[\{SU3A[i], SU3[j], SU3[a]\}]] \mmaDef{S1}[\{SU3[a]\}]\},
      \{0,0\},
      \{gp/3 \mmaDef{Bar}[\mmaDef{S1}[\{SU3[j]\}]], -gp/3 \mmaDef{S1}[\{SU3[j]\}]\}
     \},
    \mmaDef{M}[\(\Phi\)] -> \{MS, MS\},
    \mmaDef{G}[\mmaDef{A}] -> \{\{SU3A\}, \{SU2A\}, \{\}\},
    \mmaDef{G}[\(\Phi\)] -> \{\{\mmaDef{Bar}@SU3, Y[1/3]\}, \mmaDef{Bar}@\{\mmaDef{Bar}@SU3, Y[1/3]\}\}
   \} 
  ]//\mmaDef{EvaluateLoopFunctions}
\end{mmaCell}
\begin{mmaCell}{Output}
  -\mmaFrac{1}{9} \mmaSup{MS}{2} (\mmaSup{gp}{2} + 9\(\,\)\mmaSup{gc}{2} \mmaSub{C}{2}[SU3]) \bigg(1 + Log\(\Big[\)\mmaFrac{\mmaSup{\(\overline{\mu}\)}{2}}{\mmaSup{MS}{2}}\Big]\bigg)\(\,\)\mmaSup{\(\overline{S1}\)}{a}\(\,\)\mmaSup{\(S1\)}{a}
\end{mmaCell}
Since we are introducing field substitutions with ``open covariant derivatives'' and with gauge indices, some comments on the notation of the input are in order:
\begin{enumerate}[i)]
    \item As previously mentioned, the $X_{SA}^{\scriptscriptstyle[4,3]}$ and $X_{AS}^{\scriptscriptstyle[4,3]}$ interactions in~\eqref{eq:S1Xterms} and~\eqref{eq:XAS} contain terms with one open covariant derivative. At present, \pkg{SuperTracer} only supports $X$ substitutions for terms with up to two open covariant derivatives, namely for $X_0$, $X_1^\mu$ and $X_2^{\mu\nu}$ in the expansion in~\eqref{eq:Xexpansion}. When providing $X$ substitutions, a separate rule for any relevant $X_0$, $X_1^\mu$ and/or $X_2^{\mu\nu}$ must be added. They all start with the replacement rule \mmaInlineCell[]{Input}{\{FieldType1,FieldType2\}->...} but they are differentiated by the open indices in the rule: no open index for $X_0$ (as we did with all substitutions so far), $\mu$ for $X_1^\mu$, and $\mu$ and $\nu$ for $X_2^{\mu\nu}$. In the present example, only the $X_1^\mu$ part of $X_{SA}^{\scriptscriptstyle[4,3]}$ and $X_{AS}^{\scriptscriptstyle[4,3]}$ contribute at dimension six, as trivially seen by adding the $X$ term dimensions in the supertrace of our example. Hence, we only need to input a replacement rule for this term. Indeed, the two substitutions in the example above contain an open $\mu$ index in \mmaInlineCell[]{Input}{\mmaDef{g}[\mmaUnd{\(\alpha\)}[j], \mmaUnd{\(\mu\)}]}. The symbols used for entering open Lorentz indices in the substitution must always be \mmaInlineCell[]{Input}{\mmaUnd{\(\mu\)}} in the case of one open index or \mmaInlineCell[]{Input}{\mmaUnd{\(\mu\)}} and \mmaInlineCell[]{Input}{\mmaUnd{\(\nu\)}} for two open indices.
    
    \item The input of gauge indices in $X$ substitutions requires a representation label defining the kind of index, specified as \mmaInlineCell[]{Input}{rep[index]}. In our example, we chose \mmaInlineCell[]{Input}{SU3A} and \mmaInlineCell[]{Input}{SU3} to distinguish adjoint and fundamental $\SU(3)_c$ indices, respectively, although any label names preferred by the user are equally valid. As in the vector-like fermion example, the action of $G_{\mu\nu}$ on each field needs to be defined by the appropriate substitution rules, the ones for \mmaInlineCell[]{Input}{\mmaDef{G}[\mmaDef{A}]} and \mmaInlineCell[]{Input}{\mmaDef{G}[\mmaDef{\(\Phi\)}]} in our example. For instance, the substitution rule \mmaInlineCell[]{Input}{\mmaDef{G}[\(\Phi\)] -> \{\{\mmaDef{Bar}@SU3, Y[1/3]\}, \mmaDef{Bar}@\{\mmaDef{Bar}@SU3, Y[1/3]\}\}} indicates that $S_1$ transforms in the antifundamental of $\SU(3)_c$ (since we consider \mmaInlineCell[]{Input}{SU3} to denote fundamental $\SU(3)_c$ indices and \mmaInlineCell[]{Input}{\mmaDef{Bar}} gives the conjugate) and has hypercharge $1/3$ (labeled by \mmaInlineCell[]{Input}{Y}), while $S_1^\ast$ transforms in the fundamental of $\SU(3)_c$ and has hypercharge $-1/3$. In the output, all indices are displayed in the same manner as superindices of the fields (and couplings) since tracking their type is straightforward in most cases. When this is not the case, these can be made explicit by evaluating the command \pkg{ShowRep[True]}, which globally turns on the printing of index types. This behavior can be deactivated again by using \pkg{ShowRep[False]}. Also looking into the \pkg{InputForm/FullForm} yields the information about the index types.
    
    \item The \emph{open} indices in the $ X $ substitutions are all identified to the program by always giving them the same index name. For the substitution of $ X_{\eta_1 \eta_2} $ all the open indices from the $ \eta_1 $ field, must be given in the form \mmaInlineCell[]{Input}{rep[i]}, whereas the open indices from $ \eta_2 $ must be entered as \mmaInlineCell[]{Input}{rep[j]}. In both cases \mmaInlineCell[]{Input}{rep} can freely be chosen to fit the index type, but the names must always be \mmaInlineCell[]{Input}{i} and \mmaInlineCell[]{Input}{j}, respectively. For the indices contracted internally in $ X$, there are no rules as to the index names. The usage of open indices is demonstrated in the example substitution, where the reader can see the use of \mmaInlineCell[]{Input}{i} and \mmaInlineCell[]{Input}{j} in the open \mmaInlineCell[]{Input}{SU3} and \mmaInlineCell[]{Input}{SU3A} indices. We add that it is unproblematic to use the same index name multiple times in this context as e.g. \mmaInlineCell[]{Input}{rep1[i]} and \mmaInlineCell[]{Input}{rep2[i]} are recognized as different indices by \pkg{SuperTracer}.
    
    \item Gauge generators $T^a_{ij}$ are specified in \pkg{SuperTracer} as \mmaInlineCell[]{Input}{\mmaDef{T}[\{repA[a],repR[i],repR[j]\}]}. These objects present some basic properties such as being traceless $T^a_{ii}=0$ and Hermitian $\overline{ T}^a_{ij}=T^a_{ji}$. The relation $T^a_{ik}\,T^a_{kj}=C_2(R)\,\delta_{ij}$, where $C_2(R)$ is the quadratic Casimir of the representation with $i,j$ indices, is also encoded in \mmaInlineCell[]{Input}{\mmaDef{T}}, along with $T^a_{ij}\,T^b_{ji}=S_2(R)\,\delta^{ab}$ for the Dynkin index $ S_2(R)$. Other group specific properties can be added on a case-by-case basis by the user, as we show in the example notebook \textit{S1LQExample.nb}.
    
    \item Finally, it is worth noting that other indices than gauge can be included in the $X$ substitutions. By default \pkg{SuperTracer} will assume all field indices are gauge when finding the action of field-strength tensors. One can, however, assign index labels to be treated as global or flavor indices by calling \rut{AddGlobalSym[rep]}, after which all indices with the label \mmaInlineCell[]{Input}{rep} will be treated as global indices.\footnote{The list of global labels can be reset with \rut{ResetGlobalSym[]}.} \mmaInlineCell[]{Input}{\mmaDef{Flavor}} is a predefined global index, and if e.g. we were to give $ S_1 $ a flavor index, we would call it with \mmaInlineCell[]{Input}{\mmaDef{S1}[\{SU3[a], \mmaDef{Flavor}[b]\}]}. In this case, we should also make sure to account for the flavor index being contracted along $ S_1 $ propagators, which can be accounted for by providing the global labels when we set the action of the field-strength tensors on the field: \mmaInlineCell[]{Input}{\mmaDef{G}[\(\Phi\)] -> \{\{\mmaDef{Bar}@SU3, Y[1/3], \mmaDef{Flavor}\}, \mmaDef{Bar}@\{\mmaDef{Bar}@SU3, Y[1/3], \mmaDef{Flavor}\}\}}. While we do not demonstrate global indices in action here, the \textit{S1LQExample.nb} notebook example provides an example of this functionality.
\end{enumerate}
More examples, including EOM substitutions for $S_1$ and some group algebra simplifications, are provided in the \textit{S1LQExample.nb} notebook example.

As noted before, the output from \pkg{SuperTracer} can be directly compared to the results found in Ref.~\cite{Gherardi:2020det}. Comparing the full operator basis requires a significant amount of manipulations of the results due to the lack of an automatic Fierz transformation routine. However, we have done partial checks and find agreement with the expressions we checked with a single exception.\footnote{We believe the authors of Ref.~\cite{Gherardi:2020det} are missing a factor of two in their matching coefficients $\mathcal C_{HB}$ and $\mathcal C_{HG}$.}

\section{Conclusions}\label{sec:conclusions}

Computing a low-energy effective Lagrangian from a given theory is a common and exceedingly mechanical task at the beginning of most studies both in and beyond the SM. Having an automated solution greatly simplifies and accelerates these initial stages and puts matching calculations into the realm of something that can be quickly realized to test ideas without having to devote large amounts of time to it. In many cases, the more interesting phenomenology arises at the loop-level, for example when studying flavor physics. Therefore, an automated solution should be able to include at least the one-loop effects. 

\pkg{SuperTracer} is an important step in this direction. It allows for the computation of functional supertraces, which is the central part of a functional matching computation, in an automated fashion. While the diagrammatic approach to matching is arguably more common, the path integral formalism holds several advantages. First and foremost, it requires no knowledge of the operator basis, circumventing the risk of missing an operator. Secondly, the formalism lends itself incredibly well to automation, something that cannot be said for the diagrammatic approach: Finding an operator basis and then constructing and computing all contributing Feynman graphs to fix their matching coefficients is a disproportionately more complicated task to automate.  Furthermore, computing the necessary prerequisites for the functional computations is almost trivial. The labor-intensive task of performing the momentum expansion and the actual computation of the one-loop effective action is then mostly done by \pkg{SuperTracer}. 

At the current stage, performing a one-loop matching computation with \pkg{SuperTracer} still requires a significant amount of human intervention. While it is true that computing the ingredients is easy, inputting them into the program is still somewhat time-consuming and requires a certain amount of care. Furthermore, the output produces an effective Lagrangian that typically needs to be manipulated to become useful in an actual physics computation. First, SuperTracer does not provide the interaction terms $X$ or the EOMs of the heavy fields, which need to be provided by the user. For this, the program would need to know the full theory Lagrangian and derive these expressions from its functional derivatives. Second, the resulting effective Lagrangian contains redundant operators, which can be reduced by the standard methods such as reduction of products of Dirac matrices, integration-by-parts identities, field redefinitions and Fierz transformations. These shortcomings will be addressed in the upcoming release of a \pkg{Mathematica} package called \pkg{MATCHETE}~\cite{MATCHETE}, which will contain \pkg{SuperTracer} at its heart. \pkg{MATCHETE} will allow the user to input a Lagrangian and specify the power-counting rules of the fields. The program will then automatically compute the one-loop EFT Lagrangian in a minimal basis.

Already in its current form, without the expected benefits from the full release of \pkg{MATCHETE}, the program presented here provides a tremendous simplification to a one-loop matching computations, paving the way for a fully automated solution. While it often remains illuminating to perform parts of these computations manually, a computer program can provide valuable cross-checks. We believe however, that the outlook of fully relegating the drawn-out task of a matching computation to a machine has exciting implications for model building and phenomenology in the future.
 
\subsection*{Acknowledgements}

We are grateful to Timothy Cohen, Xiaochuan Lu, and Zhengkang Zhang for communications about their related work~\cite{Cohen:2020qvb}, for providing cross-checks for our program and for coordinating the release date of their package with ours. 
The work of J.F. was supported by the Cluster of Excellence `Precision Physics, Fundamental Interactions, and Structure of Matter (PRISMA+ EXC 2118/1) funded by the German Research Foundation (DFG) within the German Excellence Strategy (Project ID 39083149). 
J.P. and F.W. have received funding from the European Research Council (ERC) under the European Union's Horizon 2020 research and innovation program under grant agreement 833280 (FLAY).
J.P., A.E.T., and F.W. have received funding by the Swiss National Science Foundation (SNF) under contract 200021-175940. 
The work of A.E.T. has received funding from the Swiss National Science Foundation (SNF) through the Eccellenza Professorial Fellowship ``Flavor Physics at the High Energy Frontier'' project number 186866.

\app

\subsection{Special SuperTracer variables}
\label{app:GlobalPro}

For completeness, Table~\ref{tab:SpecialVariables} provides a list of all public \pkg{SuperTracer} variables that were not described in Section~\ref{sec:SuperTracer}. These variables are used internally in \pkg{SuperTracer} outputs, as can be seen explicitly by applying to them the \pkg{Mathematica} routine \pkg{InputForm/FullForm}. While knowing the internal representation of symbols is useful for further manipulations of \pkg{SuperTracer} outputs, the output is formatted to make it as intuitive as possible for the user.  
\begin{center}
    \renewcommand{\arraystretch}{1.2}
\begin{longtable}{l p{.5\textwidth}}

    \textbf{Internal notation} & \textbf{Description}  \\
    \toprule
    
    \mmaInlineCell[]{Input}{\mmaDef{LTerm}[cof, op]} & Denotes a Lagrangian operator \mmaInlineCell[]{Input}{op} with coefficient \mmaInlineCell[]{Input}{cof}.~\medskip \\
    
    \mmaInlineCell[]{Input}{\mmaDef{li}[seq]} & Denotes a sequence \mmaInlineCell[]{Input}{seq} of Lorentz indices.~\medskip\\
    
    \mmaInlineCell[]{Input}{\mmaDef{DiracProduct}[seq]}\hspace{6cm} & Represents the product of Dirac matrices, charge conjugation matrices and chiral projectors in the sequence \mmaInlineCell[]{Input}{seq}. Argument \mmaInlineCell[]{Input}{\mmaDef{li}[\(\mu\)]} is used for $ \gamma_\mu$, \mmaInlineCell[]{Input}{\mmaDef{li}[\(\mu, ...\)]} for $ \Gamma^{\scriptscriptstyle (n)}_{\mu,\ldots} $, and \mmaInlineCell[]{Input}{5} for $\gamma_5 $.
    The \mmaInlineCell[]{Input}{\mmaDef{DiracProduct}} head is not show in the standard output.~\medskip\\
    
    \mmaInlineCell[]{Input}{\mmaDef{Proj}[\(\pm\)1]} & Chiral projector ($+1$ for $P_R$ and $-1$ for $P_L$). It can only be used inside \mmaInlineCell[]{Input}{\mmaDef{DiracProduct}}. Its output has the same special form as \mmaInlineCell[]{Input}{\mmaDef{PL}, \mmaDef{PR}}.~\medskip\\
    
    \mmaInlineCell[]{Input}{\mmaDef{CovD}[\mmaDef{li}[\mmaUnd{<seq1>}],expr,\mmaDef{li}[\mmaUnd{<seq2>}]]} & Internal representation of a sequence of covariant derivatives, given by the optional argument \mmaInlineCell[]{Input}{\mmaUnd{<seq1>}}, acting of the expression \mmaInlineCell[]{Input}{\mmaUnd{expr}}. The expression can have Lorentz indices, given by the optional sequence \mmaInlineCell[]{Input}{\mmaUnd{<seq2>}}. This variable has a special output format, e.g. \mmaInlineCell[]{Input}{\mmaDef{CovD}[\mmaDef{li}[\mmaUnd{\(\mu\)}],f,\mmaDef{li}[]]} shows as \mmaInlineCell[]{Output}{\mmaSub{D}{\(\mu\)}f} and \mmaInlineCell[]{Input}{\mmaDef{CovD}[\mmaDef{li}[\mmaUnd{\(\mu\)},\mmaUnd{\(\nu\)}],f,\mmaDef{li}[\mmaUnd{\(\mu\)}]]} as \mmaInlineCell[]{Output}{\mmaSub{D}{\(\mu\)}\mmaSub{D}{\(\nu\)}\mmaSup{f}{\(\mu\)}}.~\medskip\\
    
    \vtop{\hbox{\strut \mmaInlineCell[]{Input}{\mmaDef{CovD}[\mmaDef{li}[<seq>],\mmaDef{Field}[label,type,}}\hbox{\strut \qquad\;\mmaInlineCell[]{Input}{indices,charge],\mmaDef{li}[]]}}} & Internal representation of a field and covariant derivatives acting on it. The optional argument \mmaInlineCell[]{Input}{\mmaUnd{<seq>}} is a sequence of Lorentz indices marking the covariant derivatives, \mmaInlineCell[]{Input}{label} is the name of the field, \mmaInlineCell[]{Input}{type} the field type (see Table~\ref{tab:MainVariables}), \mmaInlineCell[]{Input}{indices} a list of field indices, and \mmaInlineCell[]{Input}{charge} a number (or a list of numbers) indicating the field charge(s) under the Abelian group(s).~\medskip\\
    
    \mmaInlineCell[]{Input}{\mmaDef{CovD}[\mmaDef{li}[\mmaUnd{<seq>}],\mmaDef{G}[gauge],\mmaDef{li}[\mmaUnd{\(\mu\)},\mmaUnd{\(\nu\)}]]} &  Internal representation of a field-strength tensor, and covariant derivatives acting on it. The optional argument \mmaInlineCell[]{Input}{\mmaUnd{<seq>}} is a sequence of Lorentz indices marking the covariant derivatives. The argument \mmaInlineCell[]{Input}{gauge} indicates the gauge structure and can be either a symbol labeling the Abelian symmetry, or two non-Abelian indices in the form \mmaInlineCell[]{Input}{\{a,b\}}. The arguments \mmaInlineCell[]{Input}{\mmaUnd{\(\mu\)}} and \mmaInlineCell[]{Input}{\mmaUnd{\(\nu\)}} are the Lorentz indices of the field-strength tensor.~\medskip\\
    
    \mmaInlineCell[]{Input}{\mmaDef{Index}[index,rep]} & A header to distinguish internally that a certain \mmaInlineCell[]{Input}{index} is an index variable and belongs to the representation \mmaInlineCell[]{Input}{rep}. Indices entered in other routines in the form \mmaInlineCell[]{Input}{rep[index]} are internally transformed to \mmaInlineCell[]{Input}{\mmaDef{Index}[index,rep]}. For the output, indices are displayed as superscripts and by default their representation is not printed. The routine \mmaInlineCell[]{Output}{ShowRep[True]} can be evaluated to globally activate displaying indices with their representation as subscript. The stardard behaviour is recovered by evaluating \mmaInlineCell[]{Output}{ShowRep[False]}.~\medskip \\ 
    
    \mmaInlineCell[]{Input}{\mmaDef{Dim}[rep]} & Dimension of the representation \mmaInlineCell[]{Input}{rep}. This variable has the special output format \mmaInlineCell[]{Output}{\mmaSub{N}{rep}}.~\medskip\\
    
    \mmaInlineCell[]{Input}{\mmaDef{S2R}[rep]} & Dynkin index of representation \mmaInlineCell[]{Input}{rep}. This variable has the special output format \mmaInlineCell[]{Output}{\mmaSub{S}{2}[rep]}.~\medskip\\
    
    \mmaInlineCell[]{Input}{\mmaDef{C2R}[rep]} & Quadratic Casimir of representation \mmaInlineCell[]{Input}{rep}. This variable has the special output format \mmaInlineCell[]{Output}{\mmaSub{C}{2}[rep]}.~\medskip\\
    
    
    \mmaInlineCell[]{Input}{\mmaDef{AddGlobalSym}[rep]} & Adds the representation \mmaInlineCell[]{Input}{rep} to the list of global symmetry representations (except for \mmaInlineCell[]{Input}{\mmaDef{Flavor}}, all representations are local by default). The routine \mmaInlineCell[]{Input}{\mmaDef{ResetGlobalSym}[]} can be used to reset the list of global symmetry representations to \mmaInlineCell[]{Input}{\{\mmaDef{Flavor}\}}.~\medskip \\
    
    \mmaInlineCell[]{Input}{\mmaDef{CanonizeIndices}[expr]} & Brings the Lorentz indices in \mmaInlineCell[]{Input}{expr} to canonical order by attempting different index relabelings. This routine is called by \mmaInlineCell[]{Input}{\mmaDef{SuperSimplify}}.~\medskip \\
    
    \mmaInlineCell[]{Input}{\mmaDef{SimplifyOutput}[expr]} & Applies integration by parts, commutator, and Jacobi identities to match \mmaInlineCell[]{Input}{expr} to a basis of operators. This routine is called by \mmaInlineCell[]{Input}{\mmaDef{SuperSimplify}}.~\medskip \\
    
    \bottomrule
    \caption{\pkg{SuperTracer} notation relevant for output manipulations.}
    \label{tab:SpecialVariables}
\end{longtable}
\end{center}
\subsection{Diagrammatic matching for the vector-like fermion example}
\label{app:VLexample}

In this appendix we shall further outline the comparison between the diagrammatic and the functional matching calculation in Section~\ref{sec:VLexmaple}. To compare the results swiftly, we choose a basis for the effective Lagrangian closest to the output of \pkg{SuperTracer}, which can be found in the ancillary \pkg{Mathematica} notebook \textit{VLfermExample.nb}. We then compute an exhaustive set of $n$-point functions to fix the couplings of the effective theory. 

Note that in the diagrammatic approach, several $n$-point functions can match to the same effective operators since one distinguishes between amplitudes with and without extra gauge fields. As an example, consider an effective operator of the form
\begin{align}
 \bar\psi \phi i\slashed D\phi \psi\,.
\end{align}
This operator contributes to both a four-point function $\psi^2\phi^2$ as well as a five-point function with an additional gauge field. Herein lies one of the advantages of the covariant functional approach, as it computes the WC of this operator directly. On the other hand, the relation of the two amplitudes by gauge invariance serves as a valuable cross-check of the diagrammatic calculation.

Integrating out the heavy vector-like fermion $\Psi$ produces two types of matching corrections: First, we obtain hard corrections to the renormalizable interactions of the soft fields in the theory, including the ones that were not present in the UV theory. We thus split the low-energy Lagrangian into a leading-power and subleading-power piece, defining
\begin{align}
\L_\EFT = \sum_i c_i\, o_i\,.
\end{align}
At leading power, we have the one-loop effective Lagrangian:
\begin{align}
\left.\L_\EFT^\mathrm{\scriptscriptstyle (1)}\right|_{M^0} =  \frac{c_\phi}{2}(\partial_\mu\phi)(\partial^\mu\phi)- \frac{c_m}{2} \phi^2 - \frac{c_\lambda}{4!}\phi^4 - \frac{c_A}{4}F_{\mu\nu}F^{\mu\nu} + c_\psi\,\bar\psi_L (i\slashed D)\psi_L\,. \label{eq:VLF_Leff_0}
\end{align}
Note that non-vanishing matching coefficients $c_\phi$, $c_A$ and $c_\psi$ imply that the light degrees of freedom are not canonically normalized. The presentation was chosen this way to emulate the output of the functional calculation more closely. In a diagrammatic computation, one would usually assume canonically normalized light fields and include hard wave-function corrections using the LSZ formula. Since the functional trace corresponds to 1PI diagrams, such corrections are not included in the matching coefficients. Instead, we treat $o_\phi$, $o_m$, $o_A$ and $o_\psi$ as independent composite operators and match them to the hard regions of the two-point functions. The result with canonically normalized fields can be recovered by performing the field redefinitions:
\begin{align}
 \phi &\to \frac{\phi}{\sqrt{1+c_\phi}} \,, &
 A_\mu &\to \frac{A_\mu}{\sqrt{1+c_A}} \,, &
 \psi &\to \frac{\psi}{\sqrt{1+c_\psi}}\,.
\end{align}
At dimension six, we define the following effective Lagrangian, this time including both the dimension-six terms from the tree-level and one-loop Lagrangian:
\begin{align}
   \left.\L_\EFT\right|_{M^2} =& 
 \frac{c_1}{2M^2}\,\phi D^4\phi 
 +\frac{c_2}{4M^2}\,F_{\mu\nu}\partial^2 F^{\mu\nu} 
 -\frac{ic_3}{2M^2} \bar\psi \{\slashed D,D^2\}P_L\psi  \nonumber\\
 &+\frac{c_4e}{M^2}F_{\nu\rho}\bar\psi\, \Gamma^{\mu\nu\rho}P_L\, D_\mu \psi 
 + \frac{c_5e}{M^2}\,(\partial_\nu F^{\mu\nu})\bar\psi \gamma_\mu P_L \psi \nonumber \\
 &+ \frac{c_6}{2M^2}\,\bar\psi \phi  \,i\slashed D \phi P_L\psi
 + \frac{c_7}{4!M^2}\,\phi^2 D^2 \phi^2 
 - \frac{c_8}{8M^2}\, F_{\mu\nu}F^{\mu\nu} \phi^2 
 + \frac{c_{9}}{6!M^2}\,\phi^6 \,,\label{eq:VLF_Leff_2}
\end{align}
where $\Gamma^{\mu\nu\rho}=\gamma^{[\mu}\gamma^\nu\gamma^{\rho]}$. The way the Lagrangian is written, we have anticipated the coefficients $c_i$ to be real in the matching example we are considering. The first line of this Lagrangian generates power-corrections to the propagators and, in the case of charged fields, power corrections to the gauge couplings.  Note that the most general basis should also include the CP-odd counterpart of $o_8$, with one field-strength tensor replaced by its dual. Due to the absence of a source of CP-violation in the UV model, we drop it right away.

To demonstrate the matching procedure, let us begin with the example of the two-point functions. Up to second order in the expansion in $p^2/M^2$ (with $p^\mu$ being the momentum of the field), these are:
\begin{align}
 \raisebox{-4.5ex}{\includegraphics[scale=.7]{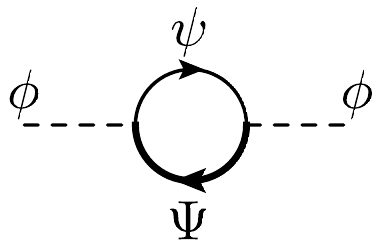}} &= \frac{i\alpha_y}{2\pi}\left\{ p^2 \left(\Delta_\mu + \frac{1}{2}\right) - 2M^2 (\Delta_\mu + 1) +\frac{p^4}{3M^2} \right\}\,, \nonumber\\
 \raisebox{-4.25ex}{\includegraphics[scale=.7]{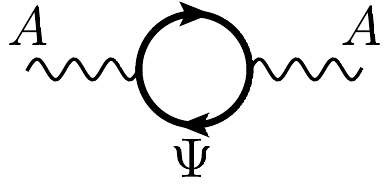}} &=  \frac{i\alpha}{3\pi} \left(p^\mu p^\nu - g^{\mu\nu}p^2 \right) \left\{\Delta_\mu + \frac{p^2}{5M^2} \right\} \,, \nonumber\\
  \raisebox{-4.5ex}{\includegraphics[scale=.7]{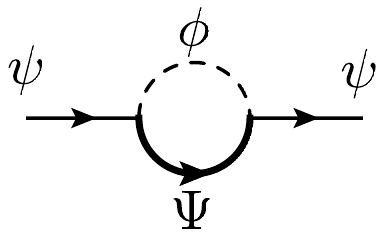}} &=  \frac{i\alpha_y}{8\pi} \slashed p P_L\left\{\left(\Delta_\mu + \frac{3}{2} \right) + \frac{2p^2}{3M^2} \right\}\,,
\end{align}
with $\Delta_\mu = 1/\epsilon + \log \mu^2/M^2$, $\alpha_y = |y|^2/4\pi$, $\alpha=e^2/4\pi$ and $d=4-2\epsilon$. The above expressions fix the ($\overline{\mathrm{MS}}$-renormalized) effective couplings:
\begin{align}
 c_\phi &= \frac{\alpha_y}{2\pi}\left( \log \frac{\mu^2}{M^2} + \frac{1}{2}\right) \,, &
 c_m    &=  \frac{\alpha_y}{\pi}M^2\left( \log \frac{\mu^2}{M^2} + 1\right) \,, &
  c_1 &= \frac{\alpha_y}{6\pi} \,, \nonumber\\
  c_A &= \frac{\alpha}{3\pi}\log \frac{\mu^2}{M^2}\,, &
c_2 &= \frac{\alpha}{15\pi}\,, &  & \nonumber\\
 c_\psi &= \frac{\alpha_y}{8\pi}\left( \log \frac{\mu^2}{M^2} + \frac{3}{2}\right)\,, &
 c_3 &= \frac{\alpha_y}{12\pi}\,.
\end{align}
Inserting these results into the Lagrangian~\eqref{eq:VLF_Leff_0} and the first line of~\eqref{eq:VLF_Leff_2} reproduces the output from the program when evaluating the following log-type and power-type supertraces
\begin{align}
 \mathrm{STr} \left\{\Delta_\Psi X^{\scriptscriptstyle[1]}_{\Psi\psi}\Delta_\psi X^{\scriptscriptstyle[1]}_{\psi\Psi} \right\}\,,\qquad 
 \mathrm{STr}\,\ln\,\Delta_\Psi\,,\qquad
 \mathrm{STr} \left\{\Delta_\Psi X^{\scriptscriptstyle[3/2]}_{\Psi\phi}\Delta_\phi X^{\scriptscriptstyle[3/2]}_{\phi\Psi} \right\} \,.
\end{align}
The operator $o_3$ is the first one in the sequence that contributes to more than one amplitude, as it is possible to contract it with up to three external photon states. Naturally, the next step is to compute the three-point function of two fermions and one gauge field. It is found to be
\begin{align}
\raisebox{-5.5ex}{\includegraphics[scale=.6]{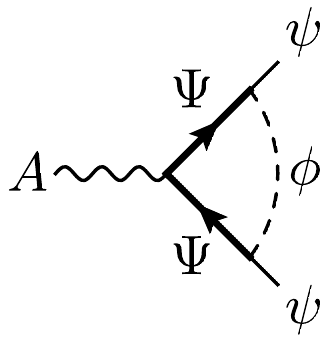}} 
= \frac{ie\alpha_y}{8\pi}\left\{ \gamma^\mu P_L \left[\Delta_\mu +\frac{3}{2}\right]  + \frac{1}{M^2}\left[
 \gamma^\mu P_L \left(\frac{11 (k_1^2+k_2^2) +7 k_1\cdot k_2}{9} \right) \right.\right.\qquad &\nonumber\\
  \left.\left. - \frac{p^\mu}{6}\slashed p P_L - \slashed k_1\gamma^\mu\slashed k_2 P_L
 \right] \right\}\,,&
\end{align}
where the fermion and anti-fermion have outgoing momenta $k_1$ and $k_2$, respectively, and we define $p^\mu = k_1^\mu -k_2^\mu$. One recognizes immediately that the first term is reproduced by $o_\psi$ with the matching condition found from the two-point function of the fermion. While in the diagrammatic calculation this is a sanity check, it never occurs in the functional calculation since the supertrace 
\[
\mathrm{STr}\left\{\Delta_\Psi X^{\scriptscriptstyle[3/2]}_{\Psi\phi}\Delta_\phi X^{\scriptscriptstyle[3/2]}_{\phi\Psi} \right\}\,,
\]
immediately gives rise to the operator $o_\psi$, generating both amplitudes. The beauty of the functional calculation shines even brighter in the subleading-power contributions. For the diagrammatic matching, one needs to first find the appropriate operator basis, derive the corresponding amplitudes, and match them to the expression above. Needless to say, this is a rather tedious exercise. On the other hand, the effective interactions are all immediately found by evaluating a single supertrace. From the diagrammatic computation we find:
\begin{align}
 c_4 &= -\frac{\alpha_y}{16\pi}\,, &
 c_5 &= \frac{\alpha_y}{9\pi}\,,
\end{align}
which matches the output from \pkg{SuperTracer} perfectly.

\begin{figure}
\centering
\includegraphics[scale=.7]{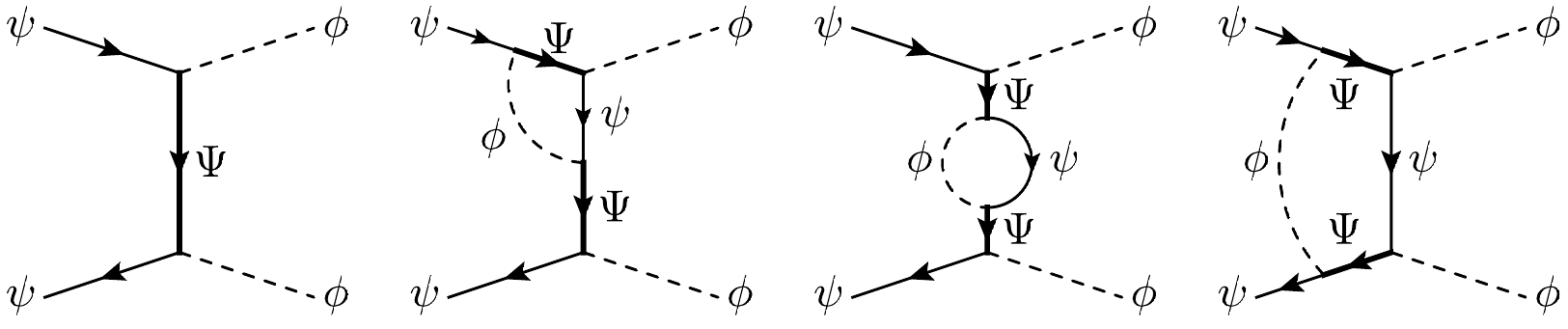}
\caption{Example Feynman diagrams contributing to the one-loop matching up to $\mathcal{O}(\alpha_y^2)$ of the effective operator $o_6$.}\label{fig:o6matching}
\end{figure}
The last example that deserves attention is the matching to the operator $o_6$. This is the first (and only) operator in $\L_\EFT^\mathrm{\scriptscriptstyle (1)}$ that is also generated at tree-level. In the diagrammatic approach, we evaluate all Feynman graphs corresponding to the amplitudes up to the desired order in the couplings. We find that the one-loop corrections proportional to $\mathcal{O}(\alpha)$ vanish in the sum over all diagrams. The non-vanishing contributions are then found from graphs akin to those shown in Fig.~\ref{fig:o6matching}. They lead us to the matching condition:
\begin{align}
 c_6 = 2|y|^2 \left[1 - \frac{\alpha_y}{2\pi}\left(\log \frac{\mu^2}{M^2}+1 \right)- \frac{\alpha_y}{8\pi}  \right]\,.
\end{align}
For the sake of comparison between the diagrammatic and the functional approach, we have split the result by diagram topology: The first term originates from the tree level graphs, the second one from the vertex graphs and the third one from the box diagrams. The propagator correction is scaleless and thus vanishes. In the functional computation, this result comes together in a somewhat different form. The supertraces yielding the vertex corrections are of the form:
\begin{align}
 \mathrm{STr}\left\{\Delta_\Psi X^{\scriptscriptstyle [1]}_{\Psi\psi} \Delta_{\psi}X^{\scriptscriptstyle [7/2]}_{\psi\phi} \Delta_\phi  X^{\scriptscriptstyle [3/2]}_{\phi\Psi} \right\} \,.
\end{align}
Note that this expression, once evaluated, involves the classical $\Psi$ through $X^{\scriptscriptstyle[7/2]}_{\psi\phi}$. To obtain the effective Lagrangian, this field has to be reduced by its equations of motion, turning the result into an expression of the form of $o_6$. Contributions corresponding to the third diagram in Fig.~\ref{fig:o6matching} never appear in the functional computation because the corresponding supertrace does not involve any heavy (quantum) fields. Finally, the box-type contributions are found directly from supertraces of the form:
\begin{align}
 \mathrm{STr}\left\{\Delta_{\Psi}\,X_{\Psi\psi}^{\scriptscriptstyle [1]}\,\Delta_\psi\,X_{\psi\Psi}^{\scriptscriptstyle [1]}\,\Delta_{\Psi}\,X_{\Psi\phi}^{\scriptscriptstyle [3/2]}\,\Delta_\phi\,X_{\phi\Psi}^{\scriptscriptstyle [3/2]}\right\}\,.
\end{align}

The rest of the calculation proceeds analogously to the concepts explained here, and we refrain from detailing every step of the computation. Instead, we simply give the remaining matching coefficients,
\begin{align}
 c_\lambda &= 24\alpha_y^2 \log \frac{\mu^2}{M^2}\,, &
 c_7 &= 13\alpha_y^2\,, &
 c_8 &= \frac{8}{3}\alpha_y\alpha\,, &
 c_9 &= \frac{15|y|^6}{\pi^2}\,,
\end{align}
which reproduces the output found from \pkg{SuperTracer} exactly.

\sectionlike{References}
\vspace{-10pt}
\nocite{apsrev41Control}
\bibliography{References}
\end{document}